\documentclass{aa}
\pdfoutput=1
\usepackage{geometry}                		

\usepackage{color}
\usepackage[varg]{txfonts}

\usepackage[dvipsnames]{xcolor}   
\usepackage{booktabs}

\usepackage{graphicx}  
\usepackage{subfig}

\newcommand{\kms}{km~s$^{-1}\,$}

\newcommand{\degm}{^{\circ}\,}

\begin{document}

   \title{The spectroscopic imprint of the pre-eruptive configuration resulting into two major coronal mass ejections}


   \author{Syntelis, P.\inst{1,2}
          \and
          Gontikakis, C.\inst{2}
          \and
          Patsourakos, S.\inst{3}
          \and
          Tsinganos, K.\inst{1}
          }

   \institute{Section of Astrophysics, Astronomy and Mechanics, Department of Physics, University of Athens, Athens, Greece, \email{psyntelis[at]phys.uoa.gr}             
         \and
           Research Center for Astronomy and Applied Mathematics, Academy of Athens, 4 Soranou Efessiou Str., Athens, Greece 
		 \and
            University of Ioannina, Department of Physics, Section of Astrogeophysics, Ioannina, Greece
             }
			
   \date{}

 
	\abstract
	{}
	{ We present a spectroscopic analysis of the pre-eruptive configuration of active region NOAA 11429, prior to two very fast  coronal mass ejections (CMEs) 
	on March 7, 2012 that are associated with this active region. We study the thermal components and the dynamics associated with the ejected flux ropes.
	}
	{Using differential emission measure (DEM) analysis of  Hinode/EIS and SDO/AIA observations, we identify the emission components of both the flux rope and the host active region.
	 We then follow the time evolution of the flux rope emission components by using AIA observations.  The plasma density and the Doppler and non-thermal velocities associated with the flux ropes are also calculated from the EIS data.
	}
	{ The eastern and western parts of the active region, in which the two different fast CMEs originated during two X-class flares, were studied separately. In both regions we identified an emission component in the temperature range of $\log T=6.8 - 7.1$ associated with the presence of flux ropes. The time evolution of the eastern region showed an increase in the mean DEM in this temperature range by an order of magnitude, 5 hours prior to the first CME. This was associated with a gradual rise and heating of the flux rope as manifested by blue-shifts and increased non-thermal velocities in \ion{Ca}{XV}~200.97\AA, respectively. An overall upward motion of the flux ropes was measured (relative blue-shifts of $\sim12$~\kms). The measured electron density was found to be $4\times 10^9 - 2 \times10^{10}$ cm$^{-3}$ (using the ratio of \ion{Ca}{XV}~181.90\AA\ over \ion{Ca}{XV}~200.97\AA). 
	We compare our findings with other works on the same AR to provide a unified picture of its evolution.
	}
	{}


   \maketitle

\section{Introduction}

Observations and models of CMEs suggest that they consist of twisted magnetic structures named flux ropes (FRs).
This has been verified by a number of limb and disk observations showing case studies of FRs associated with CMEs. 
In addition, according to the standard eruptive flare
model  \citep[CSHKP,][]{Carmichael_1964,Sturrock_1966,Hirayama_1974,Knopp_Pneuman_1976}, the ejected flux rope should contain hot multi-million degree plasma, resulting from 
plasma heating in a current sheet underneath the ejected flux.
This prediction was largely confirmed by
recent observations of the Atmospheric Imaging Assembly \citep[AIA;][]{Lemen_etal2012} on board the Solar Dynamics Observatory (SDO) \citep{Pesnell_etal2012}.  These AIA, mainly limb observations,
have demonstrated the presence of a hot FR inside the CMEs core that is either pre-existing \citep[e.g.][]{Patsourakos_etal2010, Vourlidas_etal2012} or formed shortly before or during the ejection \citep[e.g.][]{Cheng_etal2011, Zhang_etal2012, Patsourakos_etal2013}. 
Another indication of hot FRs  are the extreme-ultraviolet (EUV) and soft x-rays (SXR) sigmoids, which are non-potential magnetic field structures observed on the solar disk \citep{Canfield_etal1999}. These sigmoidal structures are in some cases associated with flux-rope formation from highly sheared field lines or in other cases are associated with pre-existing flux ropes.  \citep[e.g.][]{Tripathi_etal2009,Liu_etal2010}. Thus, it seems that a close correspondence exists between flux ropes and sigmoids.
Until now, large-scale flux-rope
statistics result only from the vast number of recorded coronographic observations. \citet{Vourlidas_etal2013} show that at least 40\% of CME coronagraphic observations in the outer corona are associated with FRs. 
Based on the statistical study of 141 events, \citet{Nindos_etal2015} show that hot flux ropes are commonly observed (49\% of their events) in both confined and eruptive major solar flares.

Following the observations, CME initiation models can also be divided into two broad categories: models with pre-existing FRs and models with FR formation during the eruption. 
Most models that form twisted magnetic structures above the photosphere in a self-consistent  manner use a buoyant sub-photospheric flux tube as initial condition. 
Such a sub-photospheric flux tube emerges partially and creates a magnetic envelope field \citep[e.g.][]{Archontis_etal2004}. 
Photospheric shearing motions cause tether-cutting of the field lines \citep[similar to][]{vanBallegooijen_Martens_1989}, creating a post-emergence flux rope above the solar surface \citep[e.g.][]{Magara_etal2001}. 
The resulting post-emergence flux ropes can lead to either confined (i.e. FR acceleration is inhibited by the envelope field) or ejective eruptions (i.e. FR escapes the simulation box). This depends on the initial conditions, such as the initial flux tube magnetic field strength, the presence of an external field and its orientation, and the shape of the flux tube (toroidal, cylindrical) \citep[e.g.][]{Manchester_etal2004,Archontis_Hood_2010,Archontis_etal2014}.
Cases of confined FR eruptions are associated with the formation of quasi-stable coronal twisted flux ropes. Eruptive cases are associated with FRs formed prior to (or during) their ejection in a CME-like manner.
During the formation of these post-emergence flux ropes, tether-cutting reconnection in the flare current sheet releases energy that heats the plasma below the flux rope and along the sheared sigmoidal structures \citep[e.g. sigmoid-to-arcade flux emergence model of][]{Archontis_etal2009}. 
The mechanisms that lead to the eruption of the coronal FR are mostly studied by models assuming a pre-existing FR positioned in the solar corona. 
For instance, the role of the torus instability 
\citep[e.g.][]{Bateman_1978,Kliem_Torok_2006,Fan_etal2007}, the kink instability \citep[e.g.][]{Sakurai_1976,Torok_Kliem_2005,Fan_2005} and the break-out mechanism \citep[e.g.][]{Antiochos_etal1999,DeVore_etal2008} in the ejection of twisted flux ropes, pre-existing
or not, has been demonstrated.
Other models of pre-existing flux ropes study the destabilization of the flux rope from equilibrium using shearing motions \citep[e.g.][]{Kusano_etal2004},
flux emergence \citep[e.g.][]{Chen_etal2000}, sympathetic eruptions \citep{Torok_etal2011}, combinations of the above, etc.
Energy release and plasma heating is expected during both the formation and the destabilization of a FR.
This suggests that confined flaring (i.e. flaring not leading to a CME) is  associated with the formation of the flux-rope structures. 
These flaring activities may have different magnitudes, all the way from ``proper'' flares down to sub-flares
\citep[e.g.][]{Patsourakos_etal2013,Tziotziou_etal2013,Chintzoglou_etal2015}, and are pertinent to all the flux rope formation mechanisms discussed above.

Most of CME observations focus on the kinematics and mechanisms of the CME eruption \citep[height-time profiles, morphology, association with flares, pre-existing flux ropes or formed during the eruption, projection effects, kinetic and magnetic energy, mass etc e.g. review of ][]{Webb_etal2012}. 
Typically, CMEs are observed by either
imagers operating in the extreme-ultraviolet and the SXRs, or by white light coronagraphs. 
There are also several  studies, using EUV observations, that   
characterize the thermal components of hot flux ropes involved in CMEs \citep[e.g.][]{Cheng_etal2012,Hannah_Kontar_2013}.
Although imaging instruments supply rapid multi-wavelength high-resolution imaging 
over wide fields of view, typically covering the entire solar disk and the inner solar
corona, they nevertheless do not supply detailed plasma information and diagnostics (density, bulk and
non-thermal velocities, etc). In addition, ambiguities and difficulties could sometimes arise when
delving into the thermal structure of the corona with imaging data  \citep[e.g.][]{Guennou_etal2012}.
These shortcomings of the imaging observations are largely taken care of by \textit{spectroscopic}
observations, which however suffer from  low cadence over
smaller fields-of-view compared
to the imaging observations.
As a result, very few detailed spectroscopic observations exist both before and during CMEs.

\citet{Gibson_etal2002} analysed a sigmoidal active region and measured electron densities and temperatures using spectral data from the Coronal Diagnostic Spectrometer \citep[CDS,][]{Harrison_etal1995} on board SOHO \citep{Domingo_etal1995}. 
Older spectroscopic measurement of pre-eruption and pre-flare plasma conditions are presented in \citet{Kundu_etal1987} using the SMM-XRP experiment. SUMER \citep{Wilhelm_etal1995} spectroscopic observations were able to measure prominence oscillations 
before its eruption \citep[e.g.][]{Chen_etal2008, Chen_2011} contributing to the understanding of pre-eruption dynamics.
Moreover, active regions before flares exhibit enhancements in the non-thermal widths of spectral lines, emitted in X-rays \citep{Antonucci_etal1984,Harra_etal2001} and  in EUV \citep{Harra_etal2009}. More  recently, \citet{Harra_etal2009} found that the non-thermal width of the  Fe XII 195\AA, recorded  by Hinode/EIS, increases before any eruptive activity starts.
In conclusion, there  are currently very few detailed spectroscopic studies of the hot plasmas in flux ropes or sigmoids, so our study  could be viewed as a contribution towards filling this important gap.

In this paper we present a detailed spectroscopic study of the pre-eruptive configuration of NOAA 11429,
thereafter, AR11429, during  March 6, 2012, using observations from SDO/AIA 
and spectroscopic observations from the EUV Imaging Spectrometer \citep[EIS;][]{Culhane_etal2007} on board Hinode \citep{Kosugi_etal2007}. 
This active region  was highly dynamic, giving rise to several confined flares and eventually to  two 
super-fast CMEs during two X-class flares. These two events were initiated in
different parts of AR11429 within an hour of each other, during the first
hour of  March 7, 2012, and were responsible for one of the most active
intervals of solar cycle 24 in terms of both solar activity and  subsequent space weather conditions \citep[e.g.][]{Patsourakos_etal2016}

First, we describe the observations used for our analysis (Section~\ref{sec:observations}). 
We then calculate the differential emission measure (DEM) of the AR using AIA and EIS observations independently. 
We identify the thermal components of the AR and the flux ropes  and follow their time evolution, 12 hours prior to the first CME 
(Section~\ref{sec:thermal_diagnostics}).
Using EIS, we calculate the plasma density, as well as the Doppler and non-thermal velocities associated with the FRs (Section~\ref{sec:dynamics}). 
Finally, we discuss our results in Section~\ref{sec:discussion}.


\begin{figure}
	\centering
	\includegraphics[width=0.9\columnwidth]{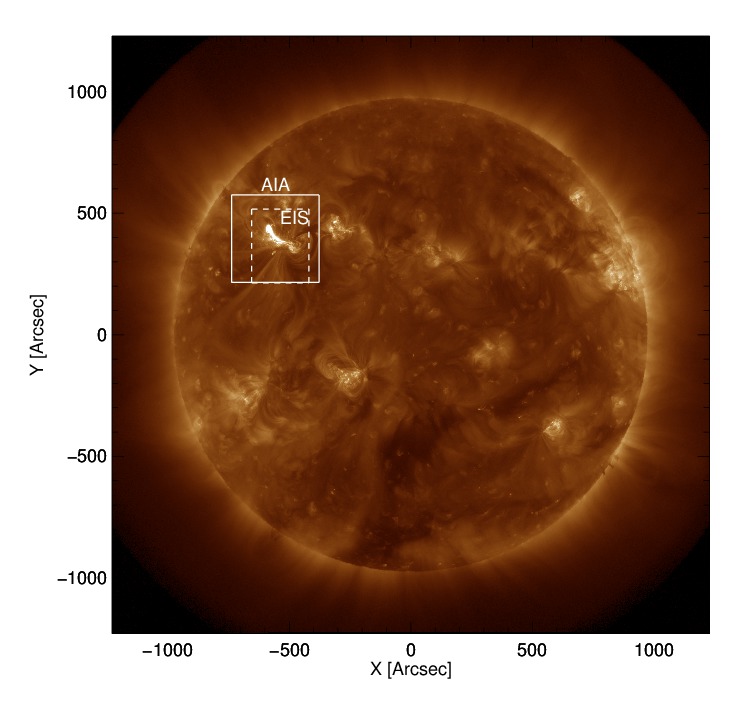}
	\caption{ Full disk AIA 94 \AA\ image on March 6, 2012, 13:25~UT. The solid lines box outlines the FoV of the AIA cutouts used in this studym and the dashed lines box indicates the FoV of the EIS rasters.
	}
	\label{fig:full_disk}
\end{figure}
\begin{figure*}
	\centering
	\includegraphics[width=0.9\textwidth]{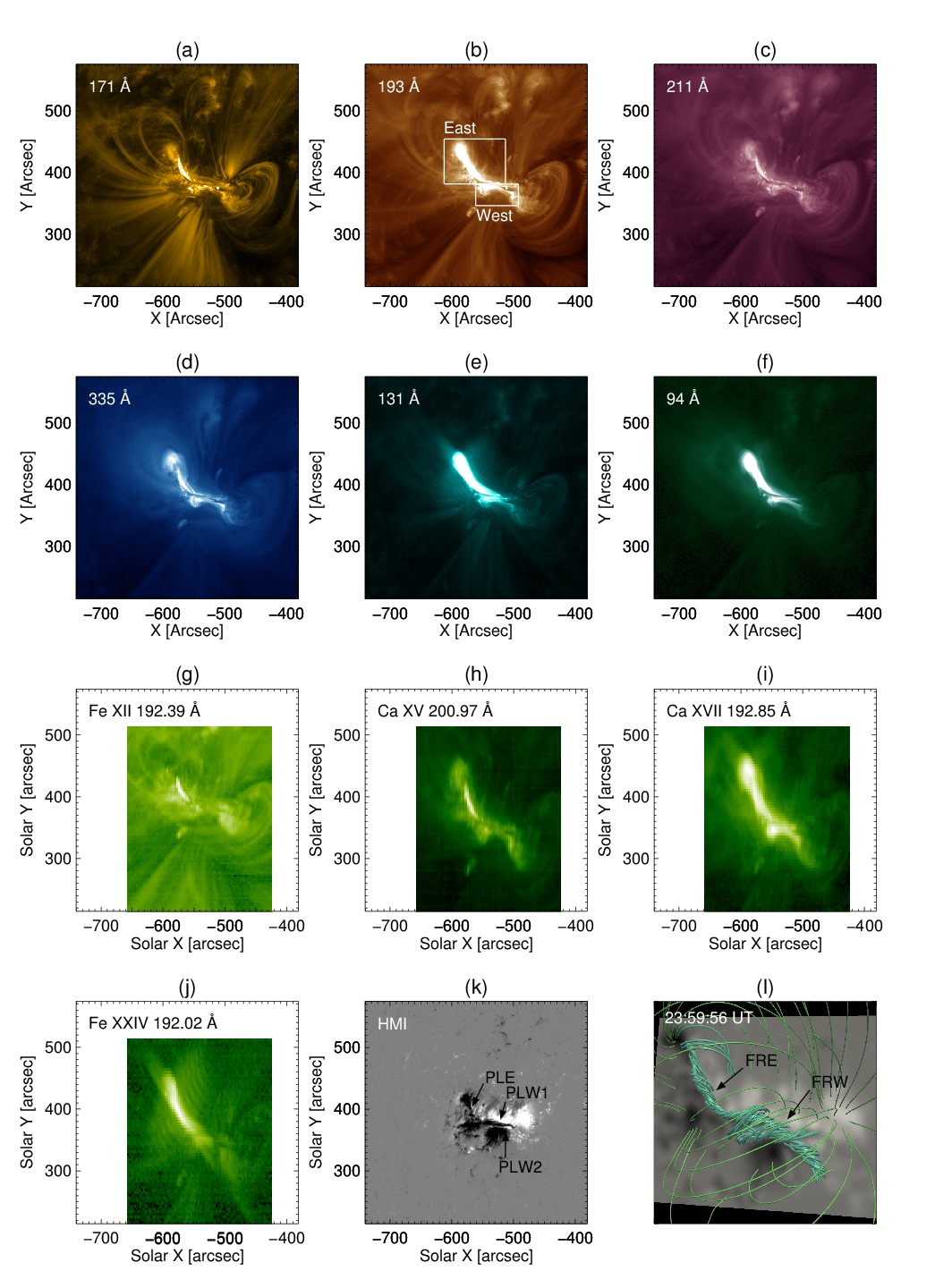}
	\caption{ Cutouts of AIA images of the AR at March 6, 12:55~UT shown in \textbf{(a)} 171\AA, \textbf{(b)} 193\AA, \textbf{(c)} 211\AA, \textbf{(d)} 335\AA, \textbf{(e)} 131\AA, \textbf{(f)} 94\AA. The white boxes in (b) define two sub-regions named East and West.
	\textbf{(g)} EIS intensity image during the 12:47-12:56~UT raster of \ion{Fe}{XII}~192.39\AA, \textbf{(h)} \ion{Ca}{XV}~200.97\AA, \textbf{(i)} \ion{Ca}{XVII}~192.85\AA\ and \textbf{(j)} \ion{Ca}{XXIV}~192.02\AA. 
	Panel sizes in EIS are the same as AIA for better comparison. White spaces in panels (g)-(j) are due to the smaller EIS FoV (see Fig~\ref{fig:full_disk}).
	\textbf{(k)} HMI line of sight magnetogram. Arrows indicate three PILs, one in the East region (PLE) and two in the West region (PLW1, PLW2). \textbf{(l)} Magnetic field extrapolation at March 6, 23:59~UT showing two flux rope structures, one in the East region (FRE) and one in the West region (FRW), surrounded by an envelope field. This figure is courtesy of G. Chintzoglou \citep[see ][]{Chintzoglou_etal2015}.
	}
	\label{fig:observations}
\end{figure*}

\section{Observations}
\label{sec:observations}

\subsection{Observational data}
Active region NOAA 11429 (boxed area in Fig. \ref{fig:full_disk}) was the origin of two CMEs early on March 7, 2012. We studied this AR twelve hours prior to the first CME using narrowband images from AIA and spectroscopic observations from EIS. 

For this study, we selected SDO/AIA observations in six coronal filters (171~\AA, 193~\AA, 211~\AA, 335~\AA, 131~\AA, 94~\AA), taken during March 6, 12:05~UT and March 7, 00:15~UT with a 10~min cadence. Data were prepared to level 1.5 using \texttt{aia\_prep.pro} and coaligned where necessary to sub-pixel accuracy. Cutouts of 359\arcsec$\times$359\arcsec~(600 by 600 pixels) were extracted from the full-disk images (Fig. \ref{fig:observations}a-f). The field of view (FoV) of the AIA cutouts is displayed as a solid box on the full-disk 193~\AA\ image of Fig. \ref{fig:full_disk}.

During the 12 hours prior to the CMEs, Hinode/EIS performed twelve raster scans of AR 11429 during two time periods: Period T1, from 12:38 to 13:31~UT, and Period T2 from 21:10 to 22:04~UT. 
The rasters had a FoV of 235\arcsec$\times$303\arcsec~(80 by 304 pixels), indicated by dashed lines in Fig.~\ref{fig:full_disk}. 
The spectral lines we analysed in this study are summarized in Table \ref{tab:lines}. Blended lines are labeled with (bl),
and lines where the blending was corrected are labeled (c). \ion{Fe}{VIII}~186.60\AA, \ion{Ca}{XIV}~193.87\AA, and \ion{S}{XIII}~256.68\AA\ were corrected according to \citet{Young_etal2007}. The \ion{Fe}{XI} 188.23\AA\ line is self-blended, but a double Gaussian fit was able to resolve both lines.
The only line that was not corrected is \ion{Fe}{XXIV}~192.02\AA, which is a flare line. It was used only in some rasters when
\ion{Fe}{XXIV}~192.02\AA\ recorded a significant signal due to M-class flares occurring in the AR.
The data were treated using \texttt{eis\_prep.pro}. 
We corrected wavelengths for slit tilts and thermal drifts. Radiometric calibration was performed according to \citet{Warren_etal2014}. Data was coaligned of the data was performed by correcting the ccd y-offset and then making further adjustments by visual inspection when necessary.
In Fig. \ref{fig:observations}g-j we plot \ion{Fe}{XII}~192.39\AA, \ion{Ca}{XV}~200.97\AA, \ion{Ca}{XVII}~192.85\AA, and \ion{Ca}{XXIV}~192.02\AA\ intensity images for comparison with the AIA panels.

\subsection{Description of AR11429}

This AR has been the subject of several studies. For instance, \citet{Chen_etal2014} studied a limb CME originating from this AR on March 3. \citet{Simoes_etal2013} studied the implosion of coronal loops on March 9. We focus on March 6-7. Then, the studied active region had two dominant bright structures in the corona. 
These structures, most prominent in the hot AIA channels at 94\AA\ and 131\AA, are located
in the eastern and the western part of the AR, respectively. They exhibit S-like shapes,
suggesting they may correspond to two flux ropes.
We name the regions where we find these structures as East and West regions (white boxes Fig.~\ref{fig:observations}b). The bright flux-rope-like structures are apparent in all AIA channels (Fig. \ref{fig:observations}a-f) and also in many EIS spectral lines (Fig.~\ref{fig:observations}h-j).

\begin{figure}
\centering
\includegraphics[width=0.95\columnwidth]{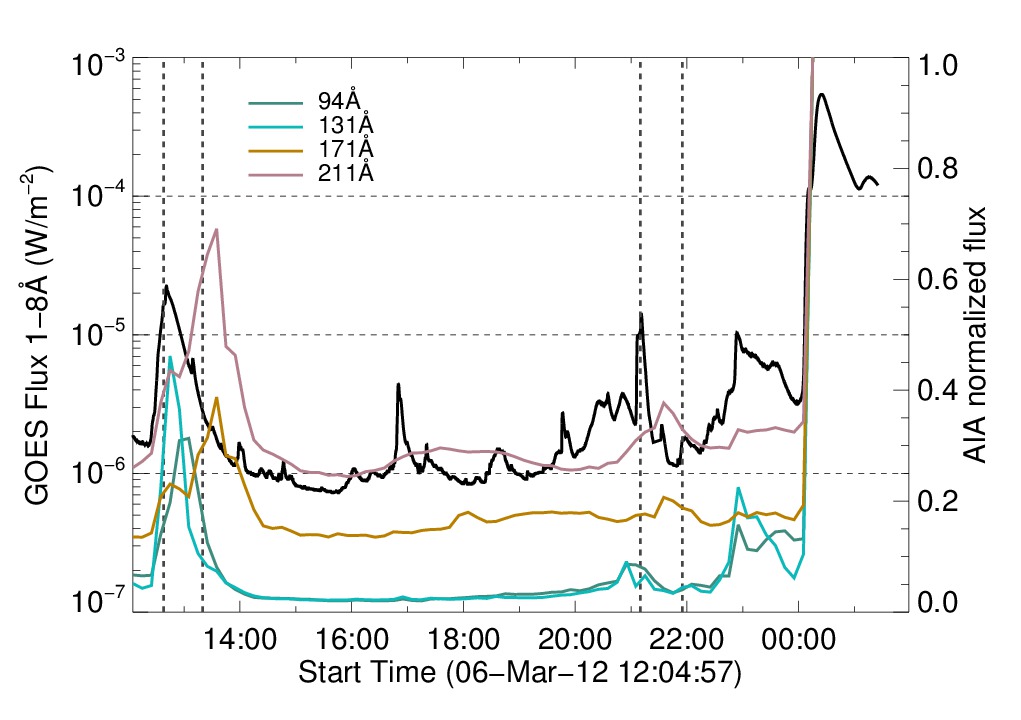}
\caption{GOES 1-8\AA~ flux from March 6, 12:05~UT to March 7, 01:25~UT. Vertical grey lines indicate the two time periods when Hinode/EIS performed rasters.
The coloured lines are the 171\AA\, 211\AA, 131\AA, and 94\AA\ AIA lightcurves taken over the whole AR cutout (white box, Fig~\ref{fig:full_disk}).
}
\label{fig:goes_fig}
\end{figure}

From line-of-sight photospheric magnetograms, recorded by the  Helioseismic and Magnetic Imager, \citep[HMI;][]{Schou_etal2012} on SDO, it is obvious that AR11429 exhibits a complex structure (e.g., Fig. \ref{fig:observations}k). Three polarity inversion lines (PILs) are found on the photosphere, and they play significant role in the evolution of the AR (arrows, Fig. \ref{fig:observations}k). One PIL is found in the East region (PLE) and two PILs, which are parallel and close to each other, are found in the West region (PLW1 and PLW2). 
The time evolution of the magnetic field during the studied 12h interval indicates that shearing motions occurs in the vicinity of these magnetic regions \citep{Wang_etal2014,Chintzoglou_etal2015}. Interactions between magnetic elements led to a series of flaring events during the studied interval, as shown by the GOES Flux in 1-8~\AA~ (Fig. \ref{fig:goes_fig}). 
The vertical grey lines indicate the periods of the Hinode/EIS rasters. 
Overplotted are the AIA 171\AA\, 211\AA, 131\AA, and 94\AA\ lightcurves. It is apparent that most of the intense recorded flarings occurred in the studied AR.
An M2.1 flare, starting 12:38~UT, coincided with the first EIS raster. This flare occurred in the East region (see Fig. \ref{fig:observations}a-g). The second set of EIS rasters  started during the peak of the M1.3 flare at 21:11~UT and extended during the decay phase of the flare. This flare occurred in the West region. All the flares during March 6 are confined, since there was no  associated CME.

During the X5.4 flare on March 7, 00:24~UT (-480$\arcsec$, 490$\arcsec$), the first CME is ejected from the East region. The second CME is ejected shortly after from the West region around March 7, 1:05~UT, during an X1.3 flare (-420$\arcsec$, 350$\arcsec$). Both of these CMEs are very fast (around 2,000 \kms) \citep{Liu_etal2013,Liu_etal2014}. The timings of the flares associated with the AR are summarized in Table \ref{tab:flares}.

Using NLFF extrapolations, \citet{Chintzoglou_etal2015} found two twisted structures in the low corona that resemble a flux rope configuration (Fig. \ref{fig:observations}l) before the eruption of the first CME on March 6, 23:59~UT. One is above PLE (named FRE) and one above both PLW1 and PLW2 (named FRW).

In the following sections, we are  going to study the thermodynamics and the plasma dynamics of the East and West regions.

\begin{table}
\centering
\caption{EIS spectral lines used in paper}
\begin{tabular}{lcc}
\toprule
Ion               & Wavelength (\AA) &  $\log$~$T$\\
\hline
Cool lines\\
\quad \ion{Si}{VII}	  & 275.36			 & 	5.80				 \\
\quad \ion{Fe}{VIII} (bl,c)	  & 186.60			 & 	5.80				 \\
Coronal and Hot lines \\
\quad \ion{Fe}{XI} 	  & 188.23			 & 	6.15				 \\
\quad \ion{Fe}{XII}	  & 192.39			 & 	6.20				 \\
\quad \ion{Fe}{XV}	  & 284.16			 & 	6.35				 \\
\quad \ion{S}{XIII} (bl,c)	  & 256.68			 & 	6.40				 \\
\quad \ion{Fe}{XVI}	  & 262.97			 & 	6.45				 \\
\quad \ion{Ca}{XIV}	  & 193.87			 & 	6.55				 \\
\quad \ion{Ca}{XV} (bl,c)	  & 200.97			 & 	6.65				 \\
\quad \ion{Ca}{XVII} (bl,c)	  & 192.85			 & 	6.75				 \\
Flare lines \\
\quad \ion{Fe}{XXIII} 	  & 263.76			 & 	7.15				 \\
\quad \ion{Fe}{XXIV} (bl) 	  & 192.02			 & 	7.25				 \\
\bottomrule
\end{tabular}
\tablefoot{Blended lines are marked with (bl) and corrected blends are marked with (c). $\log$~$T$ is the logarithm of the formation temperature of the ion. They correspond to the peaks of the curves in 
Figure~\ref{fig:goft}.}
\label{tab:lines}
\end{table}
\begin{table}
\centering
\caption{Summary of the  M and X flares occurring in the AR}
\begin{tabular}{lcc}
\toprule
 Flare         	    & Time (UT)      &  Region		\\
\hline
 M 2.1 		&		March 6, 12:38		&	East \\
 M 1.3 		&		March 6, 21:11		&	West \\
 X 5.4 		&		March 7, 00:34		&	East \\
 X 1.3 		&		March 7, 01:05		&	West \\
\bottomrule
\end{tabular}
\label{tab:flares}
\end{table}

\section{Thermal diagnostics}
\label{sec:thermal_diagnostics}

\begin{figure}
\centering
\includegraphics[width=0.95\columnwidth]{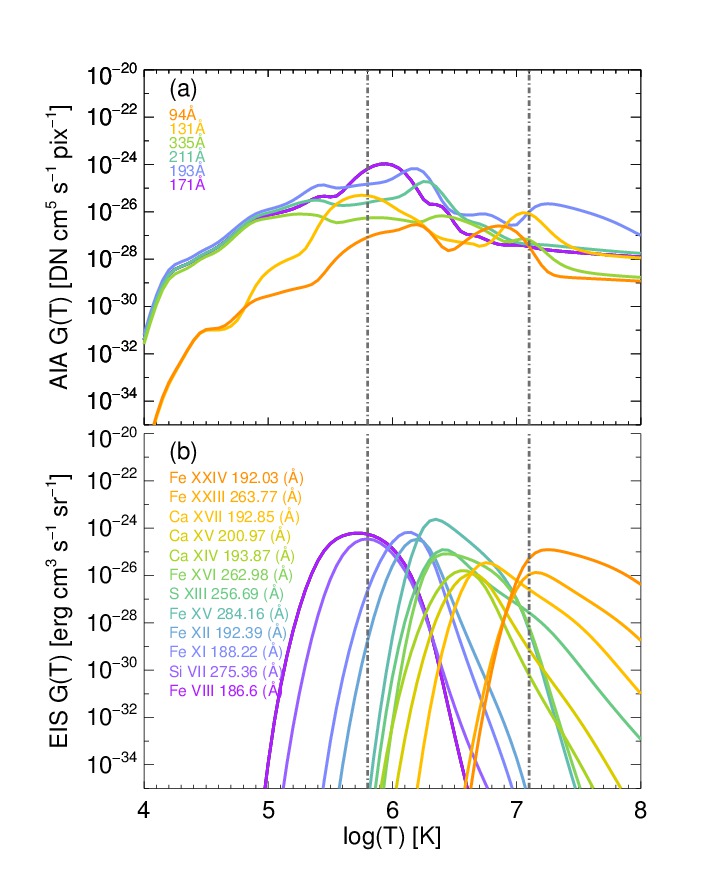}
\caption{ \textbf{(a)} Response functions of AIA channels. \textbf{(b)} Contribution functions of EIS spectral lines used in our study from CHIANTI. Grey lines indicate the temperature range used for the DEM calculations  }
\label{fig:goft}
\end{figure}

To study the thermal properties of the AR prior to the CME eruptions we perform a DEM analysis.
We calculate the DEMs for both AIA and EIS datasets. 
The two instruments produce two very different datasets. 
AIA is an imager with high spatial resolution (pixel size $\sim$0.6\arcsec) and high cadence (12~s). 
EIS performed rasters in several spectral lines with a 5~s exposure at each slit position,  using the 2\arcsec slit with scanning step size of 3\arcsec.
In addition, AIA uses narrow-band channels, each of them including a large number of spectral lines. Therefore each AIA channel is sensitive to an extended range of plasma temperatures ( Fig. \ref{fig:goft}a). 
On the other hand, EIS has greater spectral purity because it records and resolves a large number of individual spectral lines. 
Each EIS spectral line is formed in a narrow plasma temperature range, indicated by the contribution functions (Fig. \ref{fig:goft}b). 
This makes EIS better at distinguishing plasmas of different temperatures. 
The latter has high significance when performing a DEM analysis.
As a result, the comparison of AIA and EIS DEMs is not straightforward.

The aim of our DEM analysis is to deduce the thermal distribution of the flux ropes and to compare the emission components found using both instruments.
Firstly, we used the EIS DEM to find the emission component associated with the flux ropes located in the East and West regions. 
Then, we compared the results with AIA DEMs from the original AIA observations to validate whether we can identify similar emission components in AIA. 
We identified the temperature range where the flux rope emission dominates and then studied the temporal evolution of this DEM component. 
To do so, we calculated AIA DEMs during a 12~h period before the eruption of the East CME.
To calculate the DEM temporal evolution, we needed a fast DEM calculation method. We therefore used the regularized inversion method of \citet{Hannah_Kontar_2012}, which provides a fast method of calculating DEMs.
 
To do the DEM comparison, we averaged the intensities over large regions in the East and West parts of the AR (e.g. Figs. \ref{fig:dems_east}d, \ref{fig:dems_west}d). The employed spatial averaging leads to an increase in the signal-to-noise ratio, which is important particularly
for the  EIS spectral lines.
We computed  the AIA temperature response functions using the  \texttt{aia\_get\_response} routine with the /chiantifix and /evenorm keywords, using CHIANTI v7.1.4 \citep{Dere_etal1997,Landi_etal2013} to take the latest corrections in the 94\AA\ and 131\AA\ channels into account (Fig.~\ref{fig:goft}a). 
The contribution functions of  the EIS spectral lines were computed using the \texttt{gofnt} routine (Fig. \ref{fig:goft}b). 
In both AIA and EIS DEM calculations, we used the extended coronal abundances of \citet{Feldman_etal1992} (\texttt{sun\_coronal\_ext}) and the ionization equilibria provided by CHIANTI (\texttt{chianti.ioneq}). We calculated DEMs 
in the  temperature range of $\log T=5.8-7.1$ (vertical dashed-dotted grey lines in Fig. \ref{fig:goft}) in bins of 0.04~$\log T$. The method's regularization tweak parameter was set to 1. For EIS, we used  the 
emission measure
loci curve as initial guess solution.
Given the multi-peaked nature of its response functions,
we did not use the loci curve as initial condition in the calculation of  the AIA DEMs.  
However, we found that when calculating AIA DEMs with the loci curve as initial guess, the results did not differ significantly, whether in the position or in the value of the DEM maxima, from
those without the loci. 
Finally, we required positive DEM solutions for both EIS and AIA.
This was preferred since EIS data rarely converged to a positive solution. For AIA a positive solution was always found without the need to force it. 

The role of the background (BG) is crucial in the DEM analysis \citep[e.g.][]{Aschwanden_etal2011},
since the line-of-sight integration in the optically-thin corona may add
DEM components that are not necessarily connected to the structures under investigation.
To estimate the background, we computed a DEM in a box close to the East and West regions and compared the two solutions rather than subtracting them.
The EIS background DEMs were calculated using the cool and coronal lines of Table \ref{tab:lines}. 
For the East and West region DEMs, the hot and flare lines had measurable signal and were also taken into account. 

Several uncertainties
are involved in   DEM calculations. Such are
uncertainties in the atomic parameters used in the calculation of the response functions \citep[e.g.][]{Judge_etal1997}, differences between various numerical inversion methods, the ill-posedness of 
the inversion process itself, the plasma filling factor along the LOS, uncertainties
in the instrumental calibration \citep[e.g.][]{Boerner_etal2012}, etc. 
Usually the systematic errors are estimated by assuming as the spectral line's intensity error a fixed percentage of the intensity of each line. A detailed discussion on the nature and effect of these uncertainties can be found in \citet{Guennou_etal2012b,Guennou_etal2012}.
We use a 20\% of the measured intensity.
The latter is a conservative estimation of the systematic error typically involved in DEM calculations that is usually used in the literature. It resembles the estimated uncertainties involved in the calculation of the response functions \citep[e.g. see details in ][]{Judge_2010,Guennou_etal2012, Hannah_Kontar_2012}. We also note that increasing the error more than 35\%  strongly affects the resulting DEM in our case (see Sect.~\ref{sec:East_region}).

Besides the error assumption of the previous paragraph, we also present solutions that take only the Poisson noise into account, i.e. the photon counting error of the intensity measurements. For the AIA intensities $I_i$ ( where $i$ is an index corresponding to the different filtergrams, 171\AA, 195\AA, 211\AA, etc.) the error $\sigma_i$ is expressed as $\sigma_{i}\,=\, \sqrt{I_i}$. For each spectral line recorded with EIS, we used the total intensities $I_t$, derived from a Gaussian
fitting over each spectral line. The total intensities are expressed as $I_t \,=\, 2\pi I_p w$, where $I_p$ is the profile peak intensity and $w$ is the line's width. The corresponding error $\sigma_t$ is expressed as $\sigma_t^2 \,=\, I_t^2 \, ( \frac{\sigma^2_{I_p}}{I_p^2}\,+\,\frac{\sigma^2_w}{w^2} )$. 
For the EIS data, the Gaussian fits were calculated using an instrumental weighting (EIS Software Note 7, P. Young 2011). These errors are smaller than the 20\% of the measured intensities, except for the cases where the measured intensities are very faint. Such cases are the high formation temperature lines recorded with EIS.
We use both error assumptions to calculate DEMs and compare these solutions to see whether the solutions change drastically. If the solutions are very different, then the results are affected from the error assumption.

\subsection{AIA and EIS DEM analysis}

\subsubsection{East region}
\label{sec:East_region}
\begin{figure*}
\centering
\includegraphics[width=\textwidth]{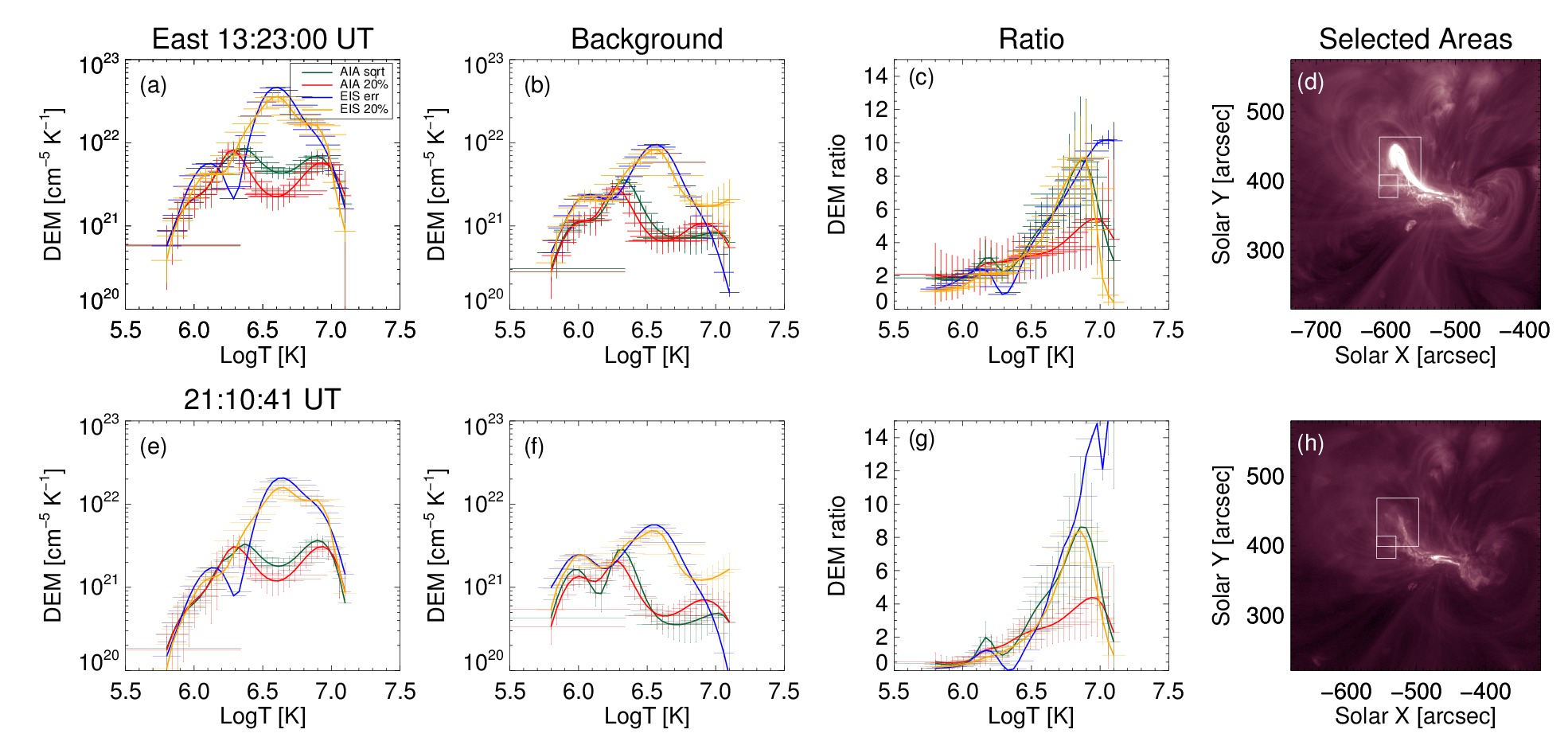}
\caption{ EIS and AIA DEMs calculated at two different times. First row is from the 13:23-13:31~UT EIS raster. Second row is from the 21:10-21:19~UT raster. The intensities used for the DEMs are average values in the East region (big box in panels (d) and (h)) and in the background (small box in panels (d) and (h)). \textbf{(a,e)} East region DEM for EIS (blue,orange) and AIA(green,red). \textbf{(b,f)} Background DEM for EIS and AIA. \textbf{(c,g)} East/background ratio. The blue and green lines use the Poisson noise as intensity error for the DEM calculation. The orange and red lines use 20\% of the intensity value as intensity error.
 \textbf{(d,h)} AIA 211\AA\ image showing the selected AR and background regions. }
\label{fig:dems_east}
\end{figure*}

\begin{figure*}
\centering
\includegraphics[width=\textwidth]{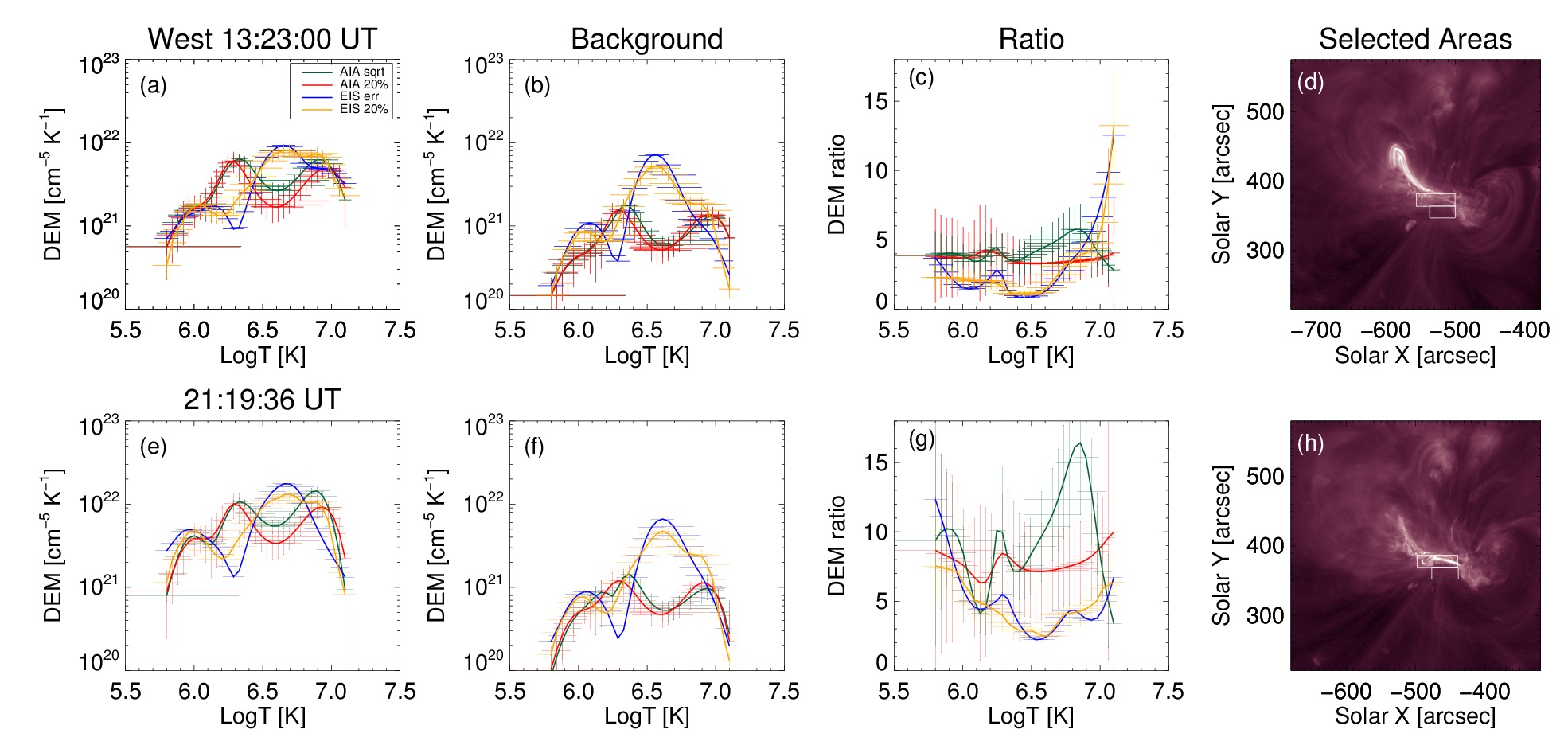}
\caption{Same as Fig.~\ref{fig:dems_east} but for the West region   }
\label{fig:dems_west}
\end{figure*}

We start with the  DEM analysis of the East region (big box Fig.~\ref{fig:dems_east}d). 
We decided to study the last raster of T1 in order to avoid the emission of the decaying M2.1 flare, which peaked at 12:38 UT (see Table~\ref{tab:flares}). In this way we instead measure the flux rope DEM (Fig.~\ref{fig:goes_fig}).
The resulting DEMs are plotted in Fig.~\ref{fig:dems_east}a. 
As mentioned before, we perform calculations assuming the intensity error (blue line, fit quality $\chi^2=2.1$) and the 20\% of the intensity (orange line, $\chi^2=1.4$).
In both error cases, we find two distinct plasma components. One low-temperature (peaking at $\log T\approx 6.1$) and one high-temperature (peaking at $\log T\approx 6.6$). For higher temperatures, the DEM curve drops 
off monotonically.

The two plasma components are also apparent (but have lower emission per temperature bin) in the BG (blue and orange lines in Fig.~\ref{fig:dems_east}b; the measurements correspond to
the  BG box of Fig.~\ref{fig:dems_east}d).
Besides the low-temperature component, we expected to have a high-temperature component in the BG. 
This happens since the background region is very close to the FR and both EIS and AIA images reveal a number of hot loops.
The ratio of the East/BG DEM
is significantly increased above $\log T= 6.5$, reaching values higher than 10. 
This indicates either excess flaring emission or the presence of a hot flux rope (or both).
For temperatures below $\log T= 6.4$, on the other hand, the ratio of the East region's DEM over the background one is lower than 4.

We then calculated  DEMs for the AIA datasets (Poisson (green) and 20\% of intensity (red) in Fig.~\ref{fig:dems_east}e).
We selected the recorded images on 13:29~UT, when EIS's slit is at  the midpoint of the East region (around $x=-580$~arcsec in Fig.~\ref{fig:dems_east}d).
In the East region, the AIA DEMs yield again two distinct peaks. A low-temperature one peaking at $\log T\approx 6.3$, and a high-temperature one peaking at around $\log T\approx 6.9$. 
Two peaks are again found in the BG and having less emission per temperature bin.

The comparison between EIS and AIA DEMs shows some similarities and some differences.
In the low-temperature regime, the two components have a comparable maximum DEM value, while the position of these two maximum values  differs by  $\log T \approx\ 0.2$.
In the high temperatures (above $\log T=6.9$ ), both DEMs drop off with temperature in the same manner.  The same applies to the low-temperature tail of the DEM.
However, in the intermediate temperatures ($\log T=6.5-6.9$), 
the EIS DEM has a peak, whereas AIA does not. 
In addition, the EIS DEM values are higher by one order of magnitude than those of AIA. On the other hand, the DEM ratios of AIA's 
East/BG yield a very similar ratio profile with EIS (Fig.~\ref{fig:dems_east}c), even for the intermediate temperatures. This similar ratio profile suggests that we do find a significantly increased emission component above 
$\log T=6.5$ (in comparison to the BG) for both instruments.
In addition, the emission components found in the intermediate temperatures, albeit different between EIS and AIA, follow the same trend in relation to the BG. 

The differences between the DEM curves, computed using  AIA and EIS  can be caused by a number of factors. As we explained in the beginning of Section~\ref{sec:thermal_diagnostics}, 
the two instruments use different observational methods (raster/image)  
producing different data samples, given that
the active region has a high time variability. 
Moreover, the intensity calibration errors 
for the two instruments
can be significant. For instance, calibrating our EIS data using different methods \citep{Warren_etal2014,Del_Zanna_2013} 
we found that intensities vary up to a factor of 2. Additionally, the EIS calibration corrections introduce different wavelength dependences that can also affect the resulting DEMs.
Finally, even if we apply the same inversion method to the data \citep{Hannah_Kontar_2012}, a 
narrowband instrument with few channels
like AIA is expected to produce a more rough DEM \citep[especially in the higher temperatures where the response function coverage is not dense,][]{Aschwanden_etal2015} in comparison to the DEM produced using EIS. The latter uses many spectral lines  and is therefore more sensitive to the details of the plasma thermal distribution.  
Therefore, it is not surprising that
the comparison between AIA and EIS shows differences in the shape of the DEM curves.

We also calculated DEMs using \texttt{xrt\_dem\_iterative2.pro} \citep[e.g. ][]{Golub_etal2007}. 
This is a forward-fitting DEM method (not an inversion one). This different method was used to check whether the differences between AIA and EIS DEMs were due to the method used or due to the nature of the data sets.
We again found a relatively good agreement for the low temperature peaks. 
A difference of $\Delta \log T\approx0.2$ was found at the location of the high-temperature peak, and there is also a difference in the value of the peak.
This suggest that the variations in the DEM curves are not solely due to the inversion method we used.
For the region/BG DEM ratio, we divided DEM functions derived from the same instrument. For this reason, the two ratios (AIA and EIS) show similar results as a function of temperature. 
At any rate, both the EIS and AIA DEMs are able to recover the two thermal components, the cool and the hot one.
We note here that both DEM calculations  are able to reproduce the measured intensities for most cases relatively well for both the Poisson  error and 20\% of the intensity error. 
In Table~\ref{tab:intensities} we present the measured intensities and the reconstructed intensities from the DEMs for the 20\% error case. 
More than half of the measured intensities, for both EIS and AIA, are reconstructed with less than a 10\% error. 
In the EIS dataset, we find differences larger
than 30\% for  \ion{Ca}{XV} and \ion{Ca}{XIV} spectral lines. For the AIA intensities, the differences are larger in the higher temperature filters (94\AA\ and 113\AA).

We calculated the same  DEM curves for the 21:10-21:19~UT raster, during the second period of EIS observations, and at t=21:17~UT for AIA.
We find again the same emission components in both the East region (Fig.~\ref{fig:dems_east}e) and the background (Fig.~\ref{fig:dems_east}f) DEM. The DEM to BG ratios also remain similar (Fig.~\ref{fig:dems_east}g).

\begin{table}
\centering
\caption{Observed and reconstructed intensities for the East DEM}
\begin{tabular}{  l  r  r  r  }
\toprule
	AIA filter / & $I_{data}$ & $I_{dem}$ & Difference (\%) \\
	EIS Spectral lines (\AA) &  &  & \\ \hline
	171 & 1504 & 1479 & -2 \\
	193 & 2883 & 2988 & 4 \\
	211 & 1398 & 1238 & -11 \\
	335 & 166 & 102 & -39 \\ 
	94 & 106 & 78 & -27 \\ 
	131 & 200 & 205 & 2 \\  \hline
	\ion{Si}{VII}~275 & 114 & 121 & 6 \\
	\ion{Fe}{VIII}~187 & 210 & 194 & -7 \\
	\ion{Fe}{XI}~188 & 1745 & 1723 & -1 \\
	\ion{Fe}{XII}~192 & 916 & 899 & -2 \\
	\ion{Fe}{XV}~284 & 27169 & 31085 & 14 \\
	\ion{S}{XIII}~257 & 2700 & 2493 & -8 \\
	\ion{Ca}{XIV}~194 & 1349 & 820 & -39 \\
	\ion{Ca}{XV}~201 & 1296 & 696 & -46 \\
	\ion{Ca}{XVII}~193 & 1878 & 2035 & 8 \\
	\ion{Fe}{XVI}~263 & 3411 & 3950 & 16 \\
	\ion{Fe}{XXIV}~192 & 128 & 129 & 0.3 \\
\bottomrule
\end{tabular}
\tablefoot{Observed and reconstructed intensities for the East DEM, during T1, with the 20\% error of Fig.~\ref{fig:dems_east}a,e. Intensities are DN/s/px for AIA and in erg/cm$^2$/s/sr for EIS.}
\label{tab:intensities}
\end{table}
\begin{figure*}
\centering
\includegraphics[width=\textwidth]{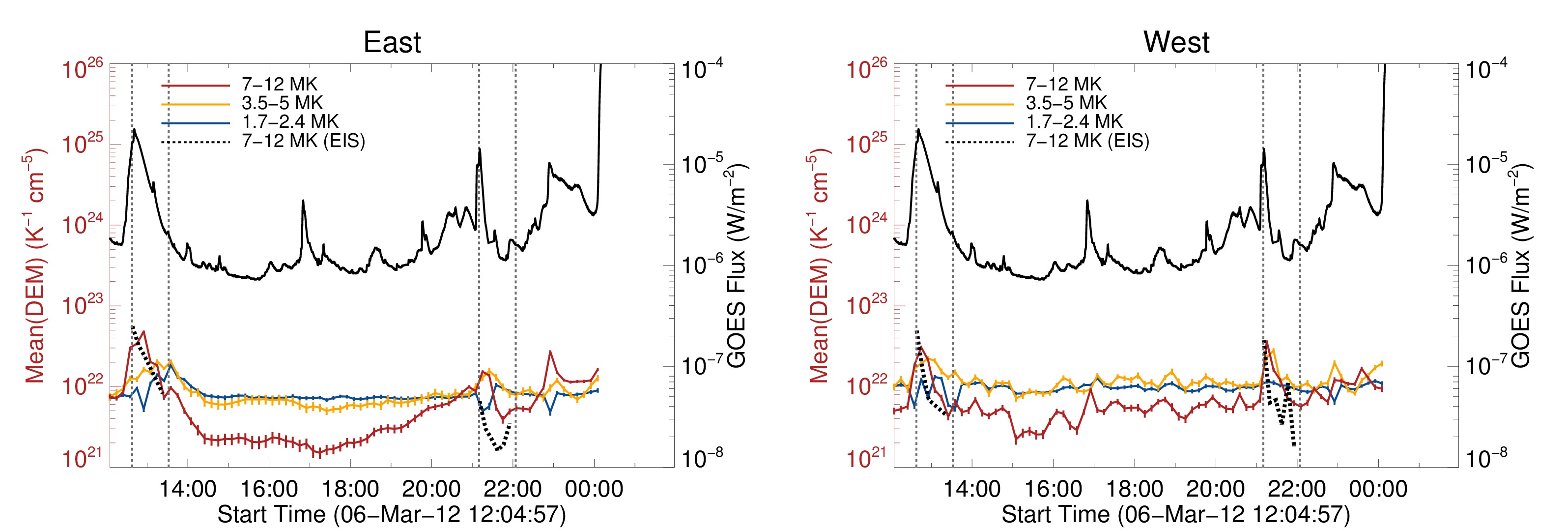}
\caption{AIA mean DEM timeseries in three temperature ranges $1.7-2.4$~MK (solid blue), $3.5-5$~MK (solid orange) and $7-12$~MK (solid red) for the \textbf{(a)} East and \textbf{(b)} West regions. 
Black dashed lines indicate the mean DEM from EIS data in the $7-12$~MK temperature range. 
The black solid lines show the GOES flux in the 1-8\AA\ channel. The vertical lines show the two time periods of the EIS rasters. }
\label{fig:dems_timeseries}
\end{figure*}

As a result, we mainly attribute the high-temperature ($\log T$ in the range 6.8-7.1) emission (and ratio) enhancement to the presence of a hot FR, which also remains hot during T2, and not to the flaring activity of the AR occurring during T1.
Again, for temperatures lower than $\log T\simeq $6.4, the ratio is always less than 4. This weak variation in the DEM between the East region and the background region indicates that the low temperature plasma component is associated with the  
background AR emission, resulting from cooler structures along the line of sight.

We note that the  DEMs from the two employed error estimates (Poisson and 20\%) result in minor morphological differences for both  T1 and T2 periods.
By trial and error, we found that we had to increase
the intensity errors  by  more than 35\% to get significant differences
in the resulting DEMs. 
In this case, the resulting EIS DEM no longer has two components. Instead, it exhibits a single peak with a maximum around $\log T=$6.6. AIA is more affected by such an error assumption. There, the resulting DEM becomes almost flat in the temperature range of $\log T=$6.2-7.1. However, the East/BG ratio increases again in high temperature.

\subsubsection{West region}
\label{sec:West_region}

We now focus on the West region. 
We selected a West region box and a background box and took the average intensity (Fig.~\ref{fig:dems_west}d).
In the 13:23-13:31~UT raster's DEM (Fig. \ref{fig:dems_west}a) we again found two plasma components when using the Poisson errors, 
but with a high $\chi^2=10.4$, making this solution unreliable. 
Using the 20\%  intensity error, 
the morphology of the DEM changes slightly, and the solution has a $\chi^2$=0.9. 
This dependence on the error suggests that the solution is not robust. 
The same applies to the background calculations (Fig. \ref{fig:dems_west}b).
The AIA DEMs also have two components found in the West region and its background (Fig. \ref{fig:dems_west}e,f). The BG component in the high temperatures is enhanced by the presence of very prominent loops found in the high temperature filters (94~\AA\ and 131~\AA). 

During T2, the West region is highly dynamic and hosts  flaring events
of various magnitudes
(see next section for more details). 
For our DEM calculations we selected the 21:19-21:28~UT raster (and 21:24~UT for AIA), 
since the 21:10-21:19~UT raster coincides with the M1.3 flare onset (see GOES curve, Fig.~\ref{fig:goes_fig}). 
We find that the emission in all temperature ranges is enhanced by a factor of 5-10 (or more) within the error bars (Fig.~\ref{fig:dems_west}g). Overall, it is unclear whether we can clearly identify a hot FR emission component in the West region owning to the region's activity and the enhanced  background. Still, the AIA ratios at 13:23~UT (Fig.~\ref{fig:dems_west}c) show indications of enhanced emission around $\log T=6.7-6.9$.

\subsection{Temporal evolution of the AIA DEM}
\label{sec:aia_timeseries}

In this section, we study the temporal evolution of two DEM components, defined as $\log T\approx$~6.2-6.4 and $\log T\approx$~6.8-7.1. For this task, we used the AIA dataset with a 10~min cadence (see Sect.~\ref{sec:observations}). We used only low-exposure data (<1~s) during flares to avoid saturated images.
Since AIA has a high signal-to-noise ratio per pixel (for all wavelengths), we computed the DEM for each pixel of the East and West regions rather than taking the average intensity of the regions (which was necessary for calculating the EIS DEMs). 
We did this because the calculation of spatially-averaged intensities
may smooth out part of the FR ``proper'' DEM due to the  possible inclusion of unrelated
regions.
The DEMs were calculated using the same method and parameters as in the previous section. 
We then produced a timeseries of DEM maps for the regions of interest (East and West, as in Figs.~\ref{fig:dems_east}d and \ref{fig:dems_west}d). 
Then and for each frame, we took the average value of the  DEMs in three temperature ranges corresponding to the AR ($\log T\approx$~6.2-6.4), the FR ($\log T\approx$~6.8-7.1), and an intermediate ($\log T\approx$~6.5-6.7) temperature range. This resulted in three mean DEM timeseries 
in these temperature ranges. 
We computed the temporal evolution of the mean DEM for both East and West regions. In Fig.~\ref{fig:dems_timeseries} we plot these results (coloured lines) and overplotted the GOES flux 
for comparison (solid black line).

We first focus on the mean DEM in the high-temperature range in the East region (Fig.~\ref{fig:dems_timeseries}a red line). We find three peaks (at 12:30~UT, 21:10~UT, and 23:15~UT), which correlate with flares recorded by GOES, signifying that the recorded flares heated the plasma to a high temperature increasing the average DEM around 7-12~MK. One interesting result from the mean 
high-temperature
DEM is an overall increase from 17:00~UT until the X5.4 flare (March 7, 00:24~UT) associated with the first CME.
The  7-12 MK component rises from $1.5 \times 10^{21}$ K$^{-1}$cm$^{-5}$  at 17:00 up to $10^{22}$ K$^{-1}$cm$^{-5}$ at 23:59~UT.

This behaviour indicates that the emission from plasmas around 7-12~MK is gradually increasing prior to the CME eruption, despite the absence of the intense flaring events (the flaring events in the East region occur only around the aforementioned peaks). 
The growth of the DEM corresponds to a gradual plasma heating and could be due to an increase in the plasma density of the FR, or an increase in the volume of plasma in this temperature range (increased $\tfrac{dh}{dT}$), or both. Such an increase is also apparent in the 94\AA\ and the 131\AA\ AIA lightcurves, albeit smaller (Fig.~\ref{fig:goes_fig}). The DEM clarifies the nature of this increase because it separates the thermal components.

In the intermediate temperature range (3.5-5~MK, orange line) we only find a slight overall increase after 17:00~UT. The most prominent features in the mid-temperatures are  emission peaks after the flare events. 
They follow the high-temperature ones by a time lag, which could be due to cooling of plasma from higher temperatures, though without excluding heating of the plasma directly into the mid-temperature range. 
The low-temperature average DEM (1.7-2.4~MK, blue line) remains overly stable, confirming our interpretation that the DEM in this temperature range is mainly the AR plasma. Few exceptions arise in some minor peaks after the flaring events that could be due to the cooling of higher temperature plasma.

\begin{figure*}
\centering
\includegraphics[width=\textwidth]{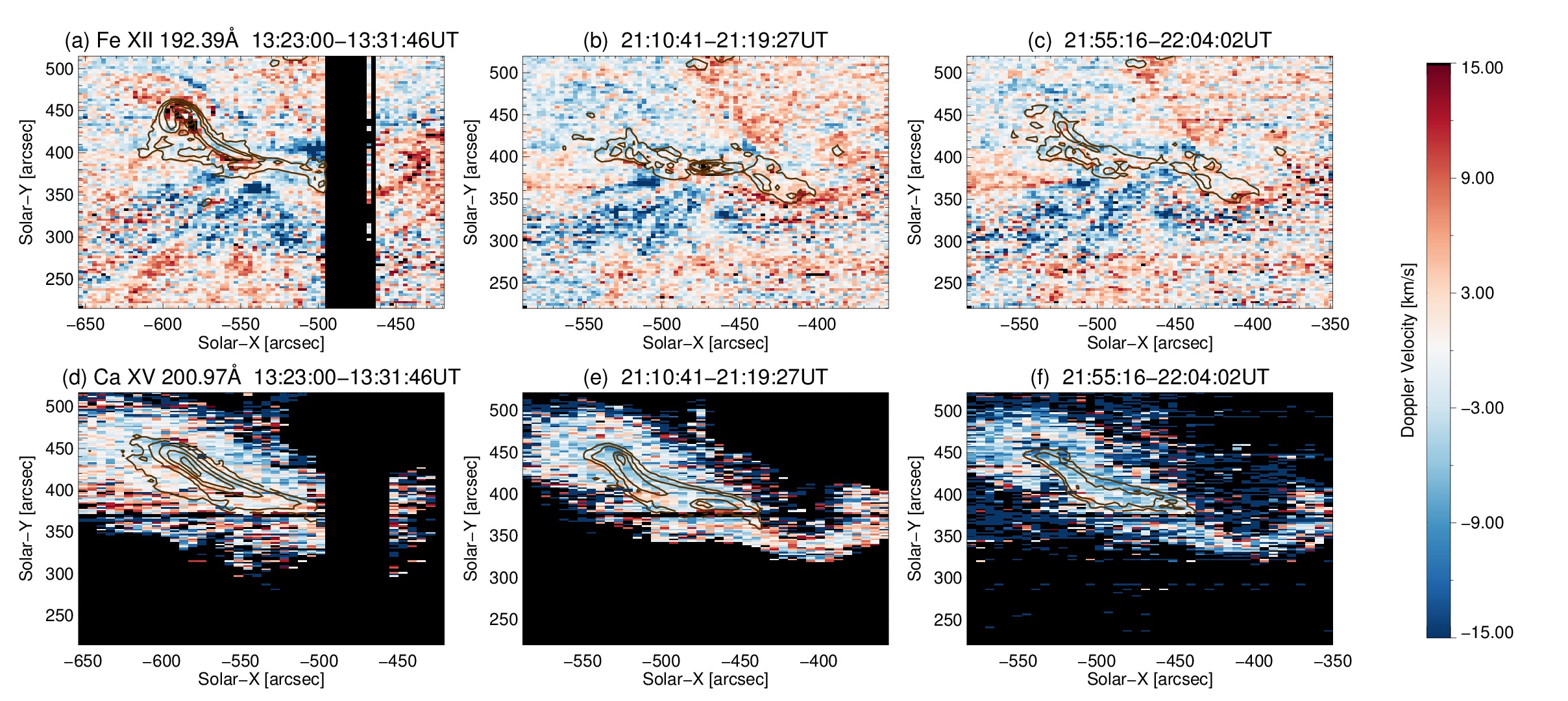}
\caption{EIS Dopplergrams for two lines during three rasters (columns). Red lines are intensity contours showing the position of the East and West FRs. First Row: \ion{Fe}{XII}~192.39\AA, Second Row: \ion{Ca}{XV}~200.97\AA. Black colour indicates missing values and bad pixels.  }
\label{fig:dopplergrams}
\end{figure*}

In the West region, for the mean DEM in the high temperature range (Fig.~\ref{fig:dems_timeseries}b, red line), we only find two of the three peaks mentioned before. 
The first peak (around 12:30~UT) is lower than the corresponding one of the East region (since the flaring occurs mainly in the East), and the second peak (around 21:10~UT) is more enhanced (since it occurs mostly in the West). The third peak is not present since this flare only occurs in the East region. 
We still find an overall increase in the mean DEM as in Fig.~\ref{fig:dems_timeseries}a, starting at 15:00~UT and continuing until 00:00~UT but with more fluctuations.
The high temperature mean DEM in the West region is highly variable,
following remarkably well the variability of the GOES  lightcurve. 
Indeed, during this 12-hour period, before the X-class flare associated with the East CME (at 12:24~UT 7-Mar), the West region exhibits a multitude of transient weak brightenings along the West FR. These
brightenings  were also recorded by GOES and do not correspond to major ($\approx$ > C2) flares.
This variability is also observed in the mid-temperature range (3.5-5~MK, orange line), 
signifying plasma cooling to this temperature domain,
 as shown by the time shifts of the red and orange peaks. We find that the low-temperature average DEM (1.7-2.4~MK, blue line) still remains overly stable. 
The observed high temporal variability may explain why the West region's DEM calculation is less robust and more sensitive to the employed errors as discussed in the Sect.~\ref{sec:West_region}.
Finally note, that the employed 10~min cadence in our DEM calculations
possibly reveals only a lower limit on the West's region variability.
We did not calculate DEMs after 00:15~UT owing to the intense flaring  that causes severe saturation to the AIA images.

The two time periods of EIS observations (vertical dashed lines, Fig.~\ref{fig:dems_timeseries}) are not big enough to capture the gradual mean DEM increase in the $7-12$~MK temperature range. Also, T1 and T2 are affected by the flaring events (notice that the AIA mean DEM drops during T1 and T2). EIS therefore cannot be used to identify the gradual mean DEM increase, but it can be used for comparison with the AIA results during T1 and T2. We plot, the mean EIS DEM in $7-12$~MK with a black dashed line. We find that in both East and West regions, the two datasets are in  relatively good agreement concerning the trend and the DEM values, supporting the validity of the AIA results.

\section{Dynamics}
\label{sec:dynamics}
\subsection{Doppler velocities}
\label{sec:doppler}

In this section we study the plasma dynamics in the AR ($\log T$ $\sim$ 6.2-6.4) and the FR ($\log T$ $\sim$ 6.8-7.1) temperature ranges using Doppler shifts.
We used \ion{Fe}{XII}~192.39 \AA\ ($\log T$=6.2),
which probes 
AR temperatures and \ion{Ca}{XV}~200.97 \AA\ ($\log T$=6.7), which
in turn probes FR temperatures. 
The EIS FoV does not include a quiet low-velocity region outside the AR to select it as a reference for the velocity calibration. 
Therefore, we calculated the reference velocity as follows. For each raster and for each spectral line, we summed all the spectral profiles from the whole AR to form a total spectral profile. This profile included the shift corrections along and across the slit (see Sect.~\ref{sec:observations}).
This spectral profile smooths out most of the noise. We then fitted this total spectral profile to find its centroid and then calculated its Doppler shift. This Doppler shift was used as a reference for our derived velocities from the EIS data. This reference value is different between different spectral lines. The differences between different rasters are not big (up to 2~\kms).
As a result, all our Doppler shift measurements are relative. 
Works on absolute velocity measurements (in AR moss \citep{Dadashi_etal2012}, on the quiet sun and AR \citep{Teriaca_etal1999}, and on the full-disk SUMER scans \citep{Peter_etal1999})
show that lines hotter than $T\approx10^6$~K are blue-shifted. \citet{Dadashi_etal2012} found that velocities drop on higher temperatures from 5km/s (1-1.5MK iron lines) to 1km/s (\ion{Fe}{XV}~284\AA, $\log$~T=6.4). 
\citet{Teriaca_etal1999} measured a 10\kms blue shift at \ion{Fe}{XII}~1242\AA ($\log$T=6.12). \citet{Peter_etal1999} measured blue shifts of 2.4\kms and 4.5\kms in \ion{Ne}{VIII}~770\AA\ ($\log$T=5.81) and \ion{Mg}{X}~625\AA\ ($\log$T=6.04), respectively.
As a result, the Doppler shifts in our datasets might be more blue-shifted than measured.

We plot the Doppler velocities in Fig.~\ref{fig:dopplergrams}. 
The first row of Fig.~\ref{fig:dopplergrams} shows the \ion{Fe}{XII}~192.39\AA\  dopplergrams (panels a,b,c), while the second row (panels d,e,f) shows the \ion{Ca}{XV}~200.97\AA\ ones.
Fig.~\ref{fig:dopplergrams} shows dopplergrams from three different rasters. 
The first column (panels a,d) shows the last raster during T1, 
while the second and third columns show the first and last rasters during T2, respectively. We over-plot intensity contours to highlight the FRs. 
Onwards, the small regions outside the main contours are not taken into account in our calculations.

\begin{figure}
\centering
\includegraphics[width=0.9\columnwidth]{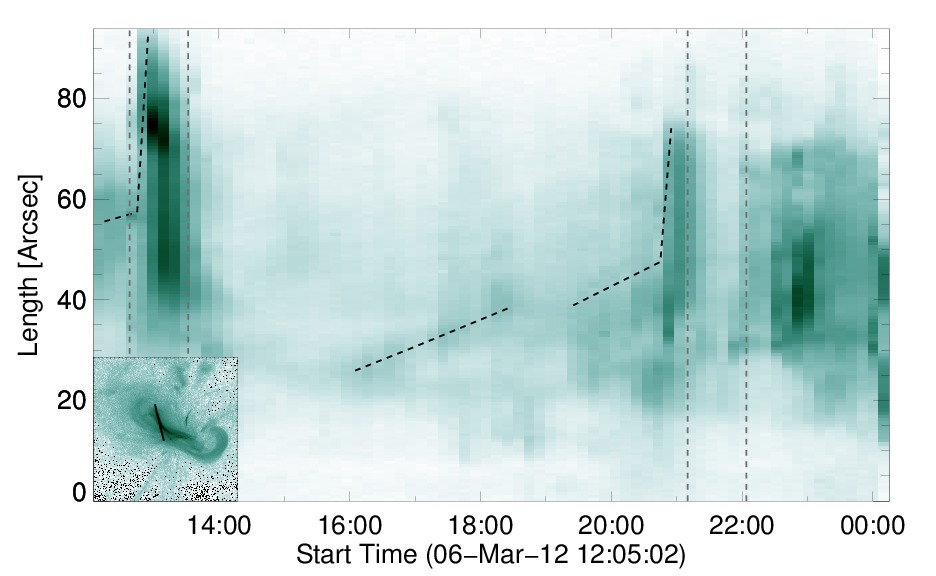}
\caption{Stack plot of AIA 94\AA\ images. The lower left sub-panel indicates the position of the used slit (black line). Vertical lines show the two time periods of the EIS rasters. Dashed black lines mark changes in slopes.}
\label{fig:jmap}
\end{figure}
\begin{figure*}
\centering
\includegraphics[width=\textwidth]{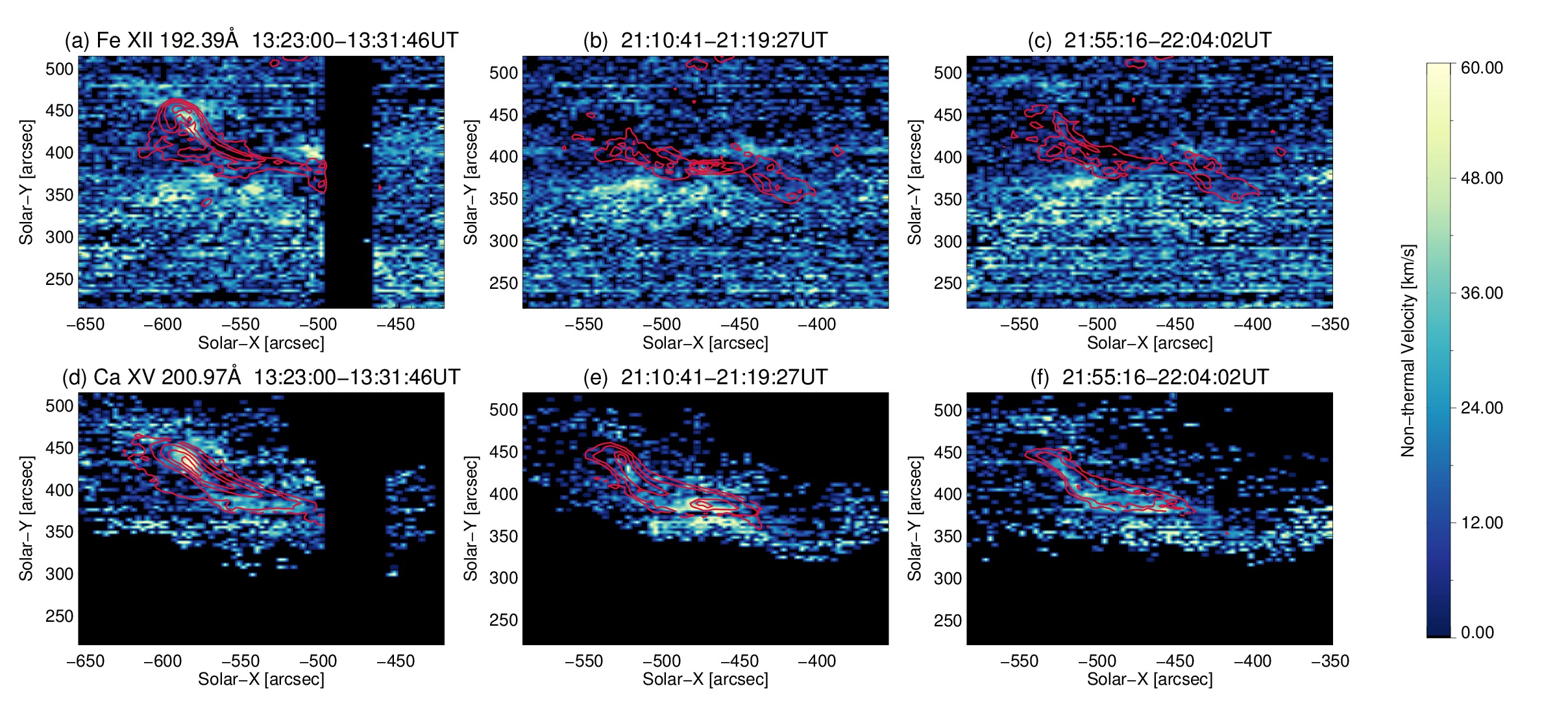}
\caption{EIS non-thermal velocity maps for two lines during three rasters (columns). Black lines are intensity contours showing the position of the East and West FRs. First Row: \ion{Fe}{XII}~192.39\AA; Second Row: \ion{Ca}{XV}~200.97\AA. Black colour indicates missing values, bad pixels, and pixels where the measured width was smaller than the instrumental.  }
\label{fig:non-thermal}
\end{figure*}

First we focus on the AR spectral line (\ion{Fe}{XII}~192.39\AA, Fig.~\ref{fig:dopplergrams}a,b,c).
Inside the FR contours, in the East region (-580\arcsec,420\arcsec, Fig.~\ref{fig:dopplergrams}a), we find upflows ($\sim-5$~\kms) in the center-most part of AR.
Downflows of around 30~\kms  are found in the periphery of the central contours. 
They indicate heating and plasma acceleration in the central part (where we have the higher intensity) and cooling associated with draining in the periphery of the structure, associated with the M2.1 flare. 
In the West region (-520\arcsec,380\arcsec), we find both smaller outflows and downflows ($\pm~5$~\kms).
During T2 we find the effects of the M1.3 flare. At  21:10-21:19~UT (Fig.~\ref{fig:dopplergrams}b), we measure higher upflows (around -10~\kms, blue colour inside contours) and downflows (around 30~\kms, red colour inside contours) . Later (21:55-22:04~UT, Fig.~\ref{fig:dopplergrams}c), the Doppler shifts decrease inside the contour ($\pm~5$~\kms), signifying that the AR probably returned to pre-flaring conditions.
During all rasters we also notice blue shifts of  $\sim$-20~\kms fanning out south of the AR core 
(around -500\arcsec, 330\arcsec).
These blue shifts could correspond to flows along closed loops that envelope the AR. 
These loops are visible in all AIA filters (see Fig.~\ref{fig:observations}). 
The AR is observed near the limb by STEREO/EUVI-B. 
These loops are apparent in all EUVI-B coronal filters 
(171\AA, 195\AA, and 284\AA) during the studied time period. 
They expand gradually, reaching up to 0.23R$_\sun$, while they become brighter during the whole pre-eruptive phase.

In the FR temperature range (\ion{Ca}{XV}~200.97\AA, Fig.~\ref{fig:dopplergrams}d,e,f), 
we find that the active region core is mostly blue-shifted.
Around the contoured FR region we measure blue shifts of -5 to -15~\kms.
We attribute these blue shifts to rise of the large scale loops we described in the previous paragraph.
Inside the FR contours, during T1 (13:23-13:32~UT raster, Fig.~\ref{fig:dopplergrams}d), we measure blue shifts around $-5$~\kms.
These upflows could stem from plasma rise or the chromospheric evaporation associated with the 12:10~UT M2.1 flare.
Downflows (around 4\kms) are again found in the periphery of the contours. 
Those downflows could be partly caused by cooling of high-temperature plasma.
During T2 (at 21:10-21:19~UT and 21:55-22:04~UT, Fig.~\ref{fig:dopplergrams}e,f),  we find increased upflows of up to $-10$~\kms 
and $-15$~\kms, respectively, inside the contours.
We note that the last EIS raster (21:55-22:04~UT, Fig.~\ref{fig:dopplergrams}f) is taken when the AR has nearly returned to non-flaring conditions. 
Therefore (at least for this raster), the \ion{Ca}{XV}~200.97\AA\ upflows are not attributed to chromospheric evaporation caused by the 21:11~UT M1.3 flare. 
Chromospheric evaporation after the flares could contribute to the upflows of previous rasters.
Our measured blue shifts are lower than other reported upflows linked to chromospheric evaporation \cite[e.g.][]{Milligan_2015}.
We thus attribute the \ion{Ca}{XV}~200.97\AA\ blue shifts mostly to rise of the FR (inside contours) and the overlying field (outside contours).

This rise in the FR (and of the overlying loops) could have a multi-phase evolution in the following manner. 
During the two EIS rastering periods, corresponding to the two confined  M-class  flares, the flares could have accelerated the FR to a new quasi-stable position. 
This could give rise to ``jumps'' in the height-time profiles during the pre-eruptive phase of a CME, as  found by, for instance, \citet{Vourlidas_etal2012} and \citet{Patsourakos_etal2013}.  
To demonstrate this complex behaviour, we plot in Fig. \ref{fig:jmap} the evolution of the FR intensity in a stack-plot format taken along the black line on the AR image in the lower left-hand corner of the figure.
We find vertical jumps (dashed black lines) during the two M-class flares. The slopes of these jumps (between 12:35-12:55~UT and 20:35-20:50~UT) indicate velocities of $\sim$15\kms, which are comparable to the measured blue shifts. In between, we find shallower  yet continuous slopes corresponding to 1-2\kms velocity. 
These swallow slopes (between 16:00 until 20:30~UT) could be an indication of a slow pre-eruptive rise of the FR \citep[similar to e.g.][]{Patsourakos_etal2010} 
possibly due to weak reconnection events beneath the FR (as the AR exhibits series of flarings), footpoint motions or similar destabilizing phenomena.
However, the AIA view point is such that these tracks could not be directly linked to height. They nevertheless indicate that the FR 
exhibits a multi-phase (non-continuous) kinematic behaviour.

\subsection{Non-thermal velocities}
\label{sec:non-thermal}

Using the same two spectral lines, we computed the non-thermal velocity $\xi$ 
from the non-thermal broadenings of the lines. For this task, 
we  first computed the line widths
from the corresponding Gaussian fittings. We then removed the instrumental width (which varies along the y-axis of the instrument slits) and  the thermal width, assuming a  temperature equal to the formation temperature of the ions emitting the observed spectral lines (see Table~\ref{tab:lines}). 
We calculated $\xi$ for the same rasters as the Doppler velocities.

We first focus on the AR spectral line (\ion{Fe}{XII}~192.39\AA, Fig. \ref{fig:non-thermal}a,b,c). 
Inside the innermost contour, we find several pixels where the instrumental width is higher than the measured one,  so that we cannot compute a non-thermal velocity (pixels in black color). 
Inside the outermost contours, non-thermal velocities of $\sim10$~\kms are measured. 
Outside the contours, the background measurements are noisy with the exception of one region located south of the  contours, where non-thermal velocities are of the order of 35~\kms. 
This region contains the large-scale loops discussed in the Doppler velocity section.
In the FR spectral line (\ion{Ca}{XV}~200.97\AA, Fig. \ref{fig:non-thermal}d,e,f), we find enhanced non-thermal velocity inside the FR contours. 
The high non-thermal velocity regions around (590\arcsec,440\arcsec) in Fig.~\ref{fig:non-thermal}d and around  (470\arcsec,380\arcsec) in Fig.~\ref{fig:non-thermal}e are found at the flare locations. 
Typical values of the non-thermal velocity of the FRs can be found outside the flare regions. For instance, inside the inner contour of Fig.~\ref{fig:non-thermal}f and along the flux rope structure, we find non-thermal velocities of around 20-40~\kms. We also find some higher values up to 60~\kms near the post-flare region. Along the flux rope and away from the flaring regions, we find similar non-thermal velocities in all the rasters. 

\begin{figure}
\centering
\includegraphics[width=0.95\columnwidth]{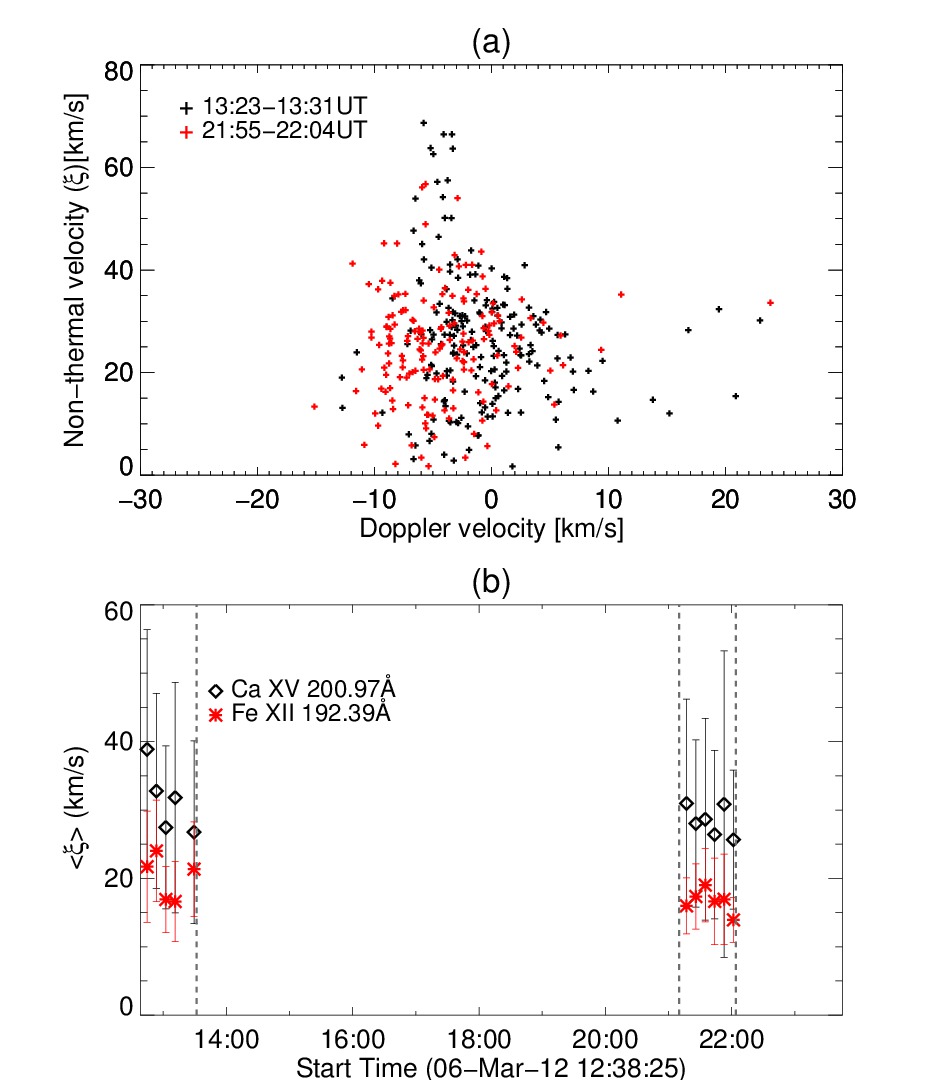}
\caption{\textbf{(a)} Non-thermal velocity in relation to Doppler flows for the \ion{Ca}{XV}~200.97\AA\ spectral line, calculated at the 13:23-13:32~UT (black) and 21:56-22:04~UT (red). All the points are pixels with intensity greater than 600 erg/cm$^2$/s/sr. \textbf{(b)} Temporal evolution of the mean non-thermal velocity inside the contours of Fig.~\ref{fig:non-thermal} for \ion{Ca}{XV}~200.97\AA\ (black) and for \ion{Fe}{XII}~192.39\AA\ (red). The error bars are the standard deviation.}
\label{fig:doppler_non-thermal}
\end{figure}

In Fig.~\ref{fig:doppler_non-thermal}a, we plot the non-thermal velocity $\xi$ 
and Doppler velocity, both computed from the \ion{Ca}{XV}~200.97\AA\ line for the 13:23-13:32~UT (black) and  the 21:55-22:04~UT (red) rasters.
These are the last rasters during T1 and T2, so they are less influenced by the two M flares. The plotted points correspond to pixels with intensities greater than 600 erg/cm$^2$/s/sr, which is the value of the outermost contour in Figs.~\ref{fig:dopplergrams},\ref{fig:non-thermal}. 
The Doppler-shift distribution of the 21:55-22:04~UT raster is shifted towards the blue relative to the Doppler shifts distribution of the 13:23-13:32~UT raster, as expected from the dopplergrams of Figs.~\ref{fig:dopplergrams}d,f. 
The difference of the average Doppler shifts between the two distributions is 4~\kms. 
During these 8~h, the AR center has moved from -37$\degm$ to -33$\degm$ longitude. 
This corresponds to a projection effect correction of the order of $1/\cos(\theta)$, which is 1.25 and 1.16 respectively and thus, assuming a radial flow, it cannot have caused the increase in the measured Doppler shift.
The non-thermal velocities range is  $\sim$ 1--70~\kms with an average value of 28~\kms and 22~\kms for T1 and T2, respectively.
The high values of $\xi$  (above 50~\kms) are found at the flares' locations.

To describe the overall evolution of the non-thermal velocities, we computed the average $\xi$ inside the contours for both AR temperatures (\ion{Fe}{XII}~192.39\AA) and FR temperatures (\ion{Ca}{XV}~200.97\AA) for all EIS observations. We plot the temporal evolution of $\xi$ in Fig.~\ref{fig:doppler_non-thermal}b.
The error bars are the standard deviation of $\xi$, describing the scatter of the data in Fig.~\ref{fig:doppler_non-thermal}a. They do not describe the non-thermal velocity errors. 
The scattering is too high to extract any meaningful temporal evolution of $\xi$. 
We find, though, that in FR temperatures the non-thermal broadening is higher than the AR one by roughly 10~\kms, 
indicating enhanced sub-resolution activities (e.g., turbulence, flows)
in hotter plasmas.

\begin{figure*}
\centering
\includegraphics[width=0.95\textwidth]{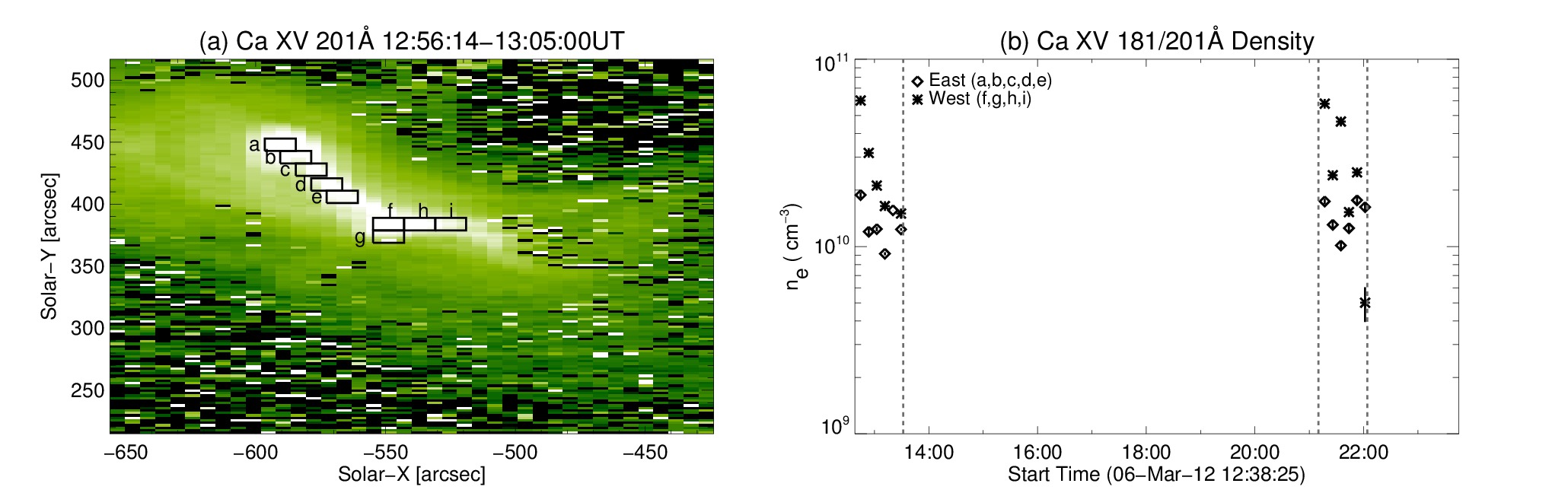}
\caption{ \textbf{(a)} Intensity map of \ion{Ca}{XV}~200.97\AA\ showing box selection along the East region (marked with a,b,c,d,e) and West region (marked with f,g,h,i). \textbf{(b)} Electron density measured from the ratio of \ion{Ca}{XV}~181.90\AA\  over \ion{Ca}{XV}~200.97\AA. Each point represents the average density along the East region boxes (diamonds) and the West region boxes (asterisk) at a certain raster.}
\label{fig:density}
\end{figure*}

\subsection{Plasma density}

To measure the plasma density around the FR temperatures, we used the density-sensitive intensity line ratio \ion{Ca}{XV}~181.90\AA/200.97\AA.
For \ion{Ca}{XV}~181.90\AA, we truncated the 181.7\AA- 182.06\AA\ spectral window for wavelengths above 182.02\AA\ to avoid the \ion{C}{IV}~182.08\AA\ line which could influence the Gaussian-fit procedure.
We used CHIANTI to derive the electron density from the observed line ratio. 
The \ion{Ca}{XV}~181.90\AA\ spectral line is weak in most of the active region except in small areas. Therefore, we selected nine box-shaped samples with a size of 4\arcsec$\times$10\arcsec. Five boxes sample the East area (named a,b,c,d,e in Fig.~\ref{fig:density}a), while the other four sample the West area of the active region (named f,g,h,i).
We also selected two regions outside the bright structure (which contains the East and West FRs) 
where we selected background profiles. These were used to correct the \ion{Ca}{XV}~200.97\AA\ intensities from the background influence. For \ion{Ca}{XV}~181.90\AA, the signal is negligible outside the AR area so that we did not apply a background correction for this line. We summed the spectral profiles in each box to increase the signal-to-noise ratio.

We then computed the line ratio and calculated the density for each of the nine boxes in Fig.~\ref{fig:density}a. We averaged the results for a,b,c,d,e to get an East region density and for f,g,h,i to get a West region density. We plot the results in Fig.~\ref{fig:density}b (asterisk for East and diamond for West).
Overall, we find that density ranges from 4$\times10^{9}$ to 5$\times10^{10}$~cm$^{-3}$. 
Since we have a few measurements, we cannot discuss temporal variations of density in detail. Only during T1 (12:47~UT to 13:13~UT) do we find a monotonic decrease in the East region's densities (from $10^{10}$~cm$^{-3}$ to $9\times10^{9}$~cm$^{-3}$), as well as in the West region's densities (from 5$\times10^{10}$~cm$^{-3}$ to 1.5$\times10^{10}$~cm$^{-3}$). We interpret this behaviour as the plasma cooling after the M2.1 flare (at 12:38~UT).

From our electron density measurements, we can separate the points that correspond to the two confined 
M flares from those corresponding to the FR by measuring the density at the location  of maximum brightness. The flare electron densities values are above  $2\times\ 10^{10}$~cm$^{-3}$. 
In contrast, the measurements taken outside the maximum brightness areas correspond to the FR. 
The FR values are in the range of $4 \times\ 10^9$ up to  $2 \times 10^{10}$ cm$^{-3}$. We also calculated the column depths  according to Young, P. (EIS solarsoft n.15 2011). 
The column depths are found from less than $4 \times\  10^7$~cm up to  $10^{10}$~cm.  For the highest electron density measurements, corresponding to flares in the West area, the column depth is as low as $6\times\ 10^7$~cm. The column depths measured outside flares, vary from $5\times\ 10^8$~cm to $10^{10}$~cm. 

\section{Discussion}
\label{sec:discussion}

\begin{table}
\centering
\caption{Summary of results}
\begin{tabular}{lcc}
\toprule
             	    & East      &  West		\\
\hline
 FR DEM ($\log T$)	& 	6.8-7.1	& 6.7-7.1	\\
 Region/BG DEM ratio 	&	10  	& 6		\\
 EIS FR peak (cm$^{-5}$~K$^{-1}$)	& 1$\times$10$^{22}$		& 1-2$\times$10$^{22}$ \\
 AIA FR peak (cm$^{-5}$~K$^{-1}$)	& 2$\times$10$^{21}$		& 5-10$\times$10$^{21}$ \\
 \hline
 Mean v$_\text{Doppler}$ (~\kms)  & \multicolumn{2}{c}{-2 to  -6}	\\
 Mean $\xi$ (~\kms)  & \multicolumn{2}{c}{22}	\\
 $n_e$ (cm$^{-3}$) & \multicolumn{2}{c}{$4\times10^9-2\times10^{10}$}\\
 Column depth (cm)  & \multicolumn{2}{c}{4$\times10^7-10^{10}$}\\
\bottomrule
\end{tabular}
\tablefoot{The lines are: the DEM range where we located the increase in the ratio of the flux rope DEM over the background DEM, the FR/BG DEM ratio's value, the EIS DEM peak at $\log$T=6.95, the AIA DEM peak at $\log$T=6.95, the mean blue-shift,  the mean non-thermal velocity, the electron density range on the FR and the column depth of the FR.} 
\label{tab:summary}
\end{table}

In this work, we have studied AR NOAA 11429, which gave rise to two major CMEs on March 7, 2012. 
We analysed two sub-regions within the AR, the East and West region. Both of them contain twisted magnetic structures (flux ropes) 
that had been found in previous studies \citep{Wang_etal2014,Chintzoglou_etal2015}.
The East and West regions are also the sites of an X5.4 flare (March 7, 00:24~UT) 
and an  X1.3 (March 7, 01:05~UT) respectively, as well as of the two CMEs associated with these flares. 
The measured properties of the flux ropes in the East and West regions are summarized in Table~\ref{tab:summary} and discussed below.

We calculated DEMs in both these regions and compared the result with the BG values using both AIA filtergrams and EIS spectroscopic observations.
The EIS DEMs showed  two  emission components  in both regions corresponding to
low ($\log T\approx$ 6.2-6.4)  and  high ($\log T\approx$ 6.5-7.1) temperatures.
The comparison to the BG showed that in the temperature range of $\log T\approx$ 6.8-7.1, the FR DEM is around 10 (East) and 6 (West) times higher.
The flux rope DEM peaks measured with EIS at  $\log T\approx$ 6.95 were approximately $1\times10^{22}$ cm$^{-5}$K$^{-1}$ (1-2$\times10^{22}$ cm$^{-5}$K$^{-1}$) for the East (West) region.
The low-temperature DEM component was attributed to the AR, since the emission component was 2 (East) and 4 (West) times higher than the BG.

We compared these results with the DEMs computed from AIA filtergrams observations.
The DEM distributions derived from EIS and AIA have some differences: the  AIA  
cooler (hotter) emission peaks were displaced by $\Delta\log T$ = 0.2  ($\Delta\log T$ = 0.3).
The region/BG ratio for both EIS and AIA DEMs exhibited  the same increasing trend  in the range $\log T\approx$6.8-7.1 (Fig.~\ref{fig:dems_east}c,g). 
The flux rope DEM peaks from AIA at  $\log T\approx$6.95 were approximately $2\times10^{21}$ cm$^{-5}$K$^{-1}$ ($5-10\times10^{21}$ cm$^{-5}$K$^{-1}$) for the East (West) region.
These values were 5 (2) times lower than the EIS DEM peaks.
One other difference was that EIS recorded enhanced emission with a peak at $\log T\approx$6.6 that was not present in the AIA DEMs.
The rough agreement between EIS and AIA DEMs
for high temperatures, suggests that the frequently used AIA DEM  determinations of hot
FR structures before and during CMEs are able to describe their emissions, at least qualitatively.
The results from our on-disk AIA DEMs are comparable to the values found in other studies that examined the DEM of CME cores at the limb. \cite{Cheng_etal2012} and \cite{Hannah_Kontar_2013} found DEM curves of similar morphology (flux rope in the range of $\log T\approx$ 6.8-7.1) and peak values (around $10^{21}$ cm$^{-5}$K$^{-1}$).

We examined the temporal evolution of the AIA's DEM components by measuring the mean DEM in the AR ($\log T\approx$~6.2-6.4), intermediate ($\log T\approx$~6.5-6.7) and the FR ($\log T\approx$~6.8-7.1) temperature ranges. The low and intermediate temperature mean DEM time series remained mostly constant (with peaks and valleys around flares). 
In the East region (which gave rise to the first CME),  we found that the mean DEM at the FR temperatures showed a gradual increase five hours prior to the CME (red line, Fig. \ref{fig:dems_timeseries}a). This gradual increase was interpreted as the combination of the gradual heating of the FR region and the gradual rise and expansion ($\tfrac{dh}{dT}$ term of DEM) of the FR magnetic field in high temperatures.

The gradual plasma rise in the  East region was inferred by the increase in blue shifts in the \ion{Ca}{XV}~200.97\AA\ line (Fig.~\ref{fig:doppler_non-thermal}a).
We associated these blue shifts with a multi-facet FR expansion/rise. During the confined M flares, the FR moves 
upwards to new quasi-stable positions similar to \citet{Vourlidas_etal2012} and \citet{Patsourakos_etal2013}. In addition, 
a phase of gradual slow rise could take place during the pre-eruptive phase of the FR with speeds lower than the Doppler velocities.
We also found that in the \ion{Ca}{XV}~200.97\AA\ spectral line the non-thermal effects were more enhanced (Fig.~\ref{fig:doppler_non-thermal}b). 
\citet{Harra_etal2009} studied the  pre-flare activity of an AR  prior to an X flare using \ion{Fe}{XII}~195\AA\ ($\log$T=6.2) and found a gradual increase in the non-thermal velocities before the flare. We did not find such a gradual increase in either of our lines (\ion{Fe}{XII}~192.39\AA, \ion{Ca}{XV}~200.97\AA) before the X flares. Such an increase might be present in other spectral lines that were not included in our non-thermal velocity study.

Using \ion{Ca}{XV}~200.97\AA\ and \ion{Ca}{XV}~181.90\AA, we measured electron densities of 4$\times10^{9}$ to 5$\times10^{10}$~cm$^{-3}$. The higher densities (above 2$\times10^{10}$~cm$^{-3}$) were  found during the  flaring events occurring in the East and West regions. The lower densities  tail (4$\times10^{9}$~cm$^{-3}$ to 2$\times10^{10}$~cm$^{-3}$) corresponds to the FR. The column depths were measured to range from 4$\times10^{8}$ to 10$^{10}$ ~cm.
Studying a sigmoidal structure observed by CDS, \citet{Gibson_etal2002} estimated the sigmoid electron density around 3$\times10^{9}$~cm$^{-3}$ (using the \ion{Si}{X} 347.40/356.05\AA\ ratio). 
During the brightening of the sigmoid, the electron density reached up to 10$^{11}$~cm$^{-3}$ (using the \ion{O}{V}/\ion{O}{IV} ratio). 
Our results are in good agreement with their measurements.  
The significant FR electron densities,
as found for the high-temperature \ion{Ca}{XV} lines, could explain
why such structures are elusive at lower temperatures.
This occurs because when a FR cools down to lower
temperatures, its density may not be sufficient to bring it
above the background.

\section{Summary}
Connecting our results to other works studying AR 11429, we present a sort summary of other results combined with ours, prior to the CME eruption in the East region.
\citet{Chintzoglou_etal2015} found that both the East and West regions contain FRs at least
12 hours before the launch of the two CMEs. 
Indeed, we identified the FR as a hot structure contributing to the DEM around $\log T\approx$ 6.8 - 7.1.
During the five hours prior to the CME eruption (after March~6, 17:00~UT) the mean DEM of this temperature range increases gradually by an order of magnitude owing to a combination of gradual heating and a gradual rise of the plasma. The rise seems to occur in a multi-phase manner, with phases of rapid rise due to confined eruptions, and phases of slow and continuous rise \citep[similar to][]{Patsourakos_etal2013}.
During this time interval, \citet{Wang_etal2014} measured an increase in the magnetic free energy of the AR (their Fig.~4) and a decrease in the negative helicity rate (which changes sign before the eruption, see their Fig.~5). 
These results are in agreement with the standard flare-CME models, demonstrating the gradual heating/gradual rise/expansion of a hot FR structure prior to its eruption, during which the magnetic energy and helicity of the system increase. Eventually the system reaches a point where it becomes unstable, leading to the CME 
and an X5.4 flare on March 7, 00:24~UT \citep[similar to works observing CMEs at the limb, e.g.][]{Patsourakos_etal2010,Vourlidas_etal2012}.
Quite possibly, this destabilization makes the adjoining West part of the AR to reconfigure its magnetic field \citep[similar to e.g.][]{Torok_etal2011}, eventually triggering there an X1.3 flare there and the CME on March 7, 01:05~UT. The propagation of the CMEs in the interplanetary medium and its effects on Earth were studied by \citet{Patsourakos_etal2016}.

\begin{acknowledgements}
The authors would like to thank the anonymous referee for the comments that helped to improve the manuscript. The authors would also like to thank Giorgos Chintzoglou for providing the extrapolation data. This research has been co-financed by the European Union (European Social Fund-ESF) and Greek national funds through the Operational Programmme ``Education and Lifelong Learning'' of the National Strategic Reference Framework (NSRF) - Research Funding Programme: Thales. Investing in knowledge society through the European Social Fund. 
We acknowledge financial support from the grant 200/790 of the research committee of the Academy of Athens.
S.P. acknowledges support from an FP7 Marie Curie
Grant (FP7-PEOPLE-2010-RG/268288). P.S acknowledges financial support from the programme Aristotelis/SIEMENS at the NOA.
\end{acknowledgements}

\bibliographystyle{aa}
\bibliography{paper}

\end{document}